\documentclass[a4paper,fleqn,table]{cas-dc}

\usepackage[numbers]{natbib}
\usepackage{amsmath,amsfonts}
\usepackage{array}
\usepackage{textcomp}
\usepackage{stfloats}
\usepackage{url}
\usepackage{verbatim}
\usepackage{soul}
\usepackage{color}
\usepackage[htt]{hyphenat}
\usepackage{multirow}
\usepackage{fontawesome}
\usepackage{booktabs}
\usepackage{graphicx, subcaption}
\usepackage[most]{tcolorbox}
\PassOptionsToPackage{hyphens}{url}
\usepackage{hyperref}
\usepackage{tikz}
\usepackage{mathtools}
\usepackage{colortbl}
\usepackage{wasysym}
\usepackage{kotex}
\usepackage{makecell}
\usepackage{xcolor}
\usepackage[linesnumbered,ruled,vlined]{algorithm2e}

\newcommand{\ourtool}{\textsc{BeaCon}}

\newcommand{\fullcercle}{$\CIRCLE$}

\newcommand{\emptycercle}{$\Circle$}

\newcommand*\circled[1]{\tikz[baseline=(char.base)]{\node[shape=circle,draw,inner sep=0.3pt] (char) {#1};}}

\newcolumntype{P}[1]{>{\centering\arraybackslash}m{#1}}

\SetKwInput{KwInput}{Require}
\SetKwFunction{FNorm}{Normal}

\begin{document}
\emergencystretch=1em

\let\WriteBookmarks\relax
\def\floatpagepagefraction{1}
\def\textpagefraction{.001}
\shorttitle{\ourtool{}: Automatic Container Policy Generation using Environment-aware Dynamic Analysis}
\shortauthors{Kang et~al.}

\title [mode = title]{\ourtool{}: Automatic Container Policy Generation using Environment-aware Dynamic Analysis}                      
\tnotemark[1]
\tnotetext[1]{This is the accepted manuscript of an article accepted for publication in \textit{Computers \& Security}.}

\author[1]{Haney Kang}[orcid=0000-0003-0866-0938]
\ead{haney1357@kaist.ac.kr}
\fnmark[1]

\author[3]{Eduard Marin}[orcid=0000-0002-5002-0187]
\ead{eduard.marinfabregas@telefonica.com}
\fnmark[1]

\author[2]{Myoungsung You}[orcid=0000-0001-5822-5243]
\ead{famous@uos.ac.kr}

\author[3]{Diego Perino}
\ead{diego.perino@gmail.com}

\author[1]{Seungwon Shin}[orcid=0000-0002-1077-5606]
\ead{claude@kaist.ac.kr}

\author[4]{Jinwoo Kim}[orcid=0000-0003-1303-8668]
\ead{jinwookim@kw.ac.kr}

\cormark[1]

\fntext[fn1]{Co-first authors}
\cortext[cor1]{Corresponding author}

\affiliation[1]{
    organization={School of Electrical Engineering, KAIST},
    addressline={291 Daehak-ro, Yuseong-gu}, 
    city={Daejeon},
    postcode={34141}, 
    state={Daejeon},
    country={Republic of Korea}}

\affiliation[2]{
    organization={School of Electrical and Computer Engineering, University of Seoul},
    addressline={163, Seoulsiripdae-ro, Dongdaemun-gu}, 
    city={Seoul},
    postcode={02504}, 
    country={Republic of Korea}}

\affiliation[3]{
    organization={Telefonica Research},
    city={Barcelona},
    country={Spain}}

\affiliation[4]{
    organization={School of Software, Kwangwoon University},
    addressline={20 Kwangwoon-ro, Nowon-gu}, 
    city={Seoul},
    postcode={01897}, 
    country={Republic of Korea}}

\begin{abstract}
This paper introduces \ourtool{}, a novel tool for the automated generation of adjustable container security policies. Unlike prior approaches, \ourtool{} leverages dynamic analysis to simulate realistic environments, uncovering container execution paths that may remain hidden during the profiling phase. To address the challenge of exploring vast profiling spaces, we employ efficient heuristics to reveal additional system events with minimal effort. In addition, \ourtool{} incorporates a security and functionality scoring mechanism to prioritize system calls and capabilities based on their impact on the host OS kernel's security and the functionality of containerized applications. By integrating these scores, \ourtool{} achieves a customized balance between security and functionality, enabling cloud providers to enforce security measures while maintaining tenant availability. We implemented a prototype of \ourtool{} using eBPF kernel technology and conducted extensive evaluations. Results from the top 15 containers, which revealed significant improvements, demonstrate that \ourtool{} identifies an average of 16.5\% additional syscalls by applying diverse environments. Furthermore, we evaluated its effectiveness in mitigating risks associated with 45 known vulnerabilities (e.g., CVEs), showcasing its potential to significantly enhance container security. Additionally, we performed proof-of-concept demonstrations for two well-known security vulnerabilities, showing that \ourtool{} successfully reduces attack surface by blocking these exploits.
\end{abstract}

\begin{highlights}
\item We design and implement \ourtool{}, a practical tool that automatically generates container policies compatible with Linux Seccomp and Capabilities.
\item We propose a novel approach for container policy generation based on emulated container environments to overcome the limitations of prior dynamic analysis techniques. We design efficient heuristics for environment exploration, an eBPF-based monitor to capture syscalls and capabilities, and security and functionality scores for intention-aware policy tuning.
\item We evaluate \ourtool{} on real-world Docker containers, demonstrating its ability to generate effective, low-overhead policies and to mitigate realistic kernel-level attacks more successfully than existing solutions.
\end{highlights}


\maketitle

\section{Introduction}

In recent years, containers have emerged as a popular choice for deploying large-scale applications in the cloud, offering increased flexibility and reduced costs. Unlike virtual machines (VMs), which require a dedicated operating system (OS) for each instance, containers on a host share the same underlying OS kernel~\cite{nist}. This shared kernel model enables faster boot times and improved resource efficiency~\cite{felter2015updated}. However, these come at the expense of providing weaker isolation compared to VMs~\cite{10.1145/3274694.3274720, reshetova2014security, felter2015updated}. Adversaries who gain control of a container (e.g., malicious tenants) can exploit vulnerabilities in the host kernel to compromise other containers or, worse, escalate privileges to take control of the host. For instance, a vulnerability in the \textit{waitid} syscall allowed attackers to perform a privilege escalation attack, escaping the container to gain access to the host~\cite{cve-2017-5123}.

One of the primary enablers of container-based attacks is that containers often operate with more privileges than they actually require, violating the principle of least privilege. By default, Docker containers have access to over 300 syscalls supported by the Linux kernel (with only 44 blocked) and are granted 14 Linux capabilities out of the 38 available~\cite{docker-default-capability}. Even if a container does not utilize some of these syscalls or capabilities, they remain accessible within the container, leaving them open to exploitation by an adversary with access to the container. This creates serious security risks, as each allowed syscall or capability effectively becomes an entry point to the kernel, which increases the chances of adversaries finding vulnerabilities in the kernel~\cite{conf/ndss/KurmusTDHRRSLK13}.

According to the NIST container security guidelines~\cite{nist}, reducing the attack surface of the host OS kernel is a promising way to alleviate the fragile isolation containers provide. Several Linux kernel mechanisms, such as \emph{Seccomp}~\cite{seccomp} and \emph{Capabilities}~\cite{capabilities}, are widely employed by cloud providers to mitigate attacks originating from malicious containers. These mechanisms enable the definition of security policies that specify the privileges granted to containers. However, these mechanisms lack the ability to automatically identify the specific privileges each container requires. To address this limitation, various solutions have been proposed that leverage dynamic analysis~\cite{lei2017speaker,Timeloops,podman,sysdig}, \emph{static analysis}~\cite{ghavamnia2020confine,Chestnut,demarinis2020sysfilter}, or a combination of the two ~\cite{blair2023automated,zhan2022shrinking}.

Dynamic analysis-based solutions~\cite{lei2017speaker,Timeloops,podman,sysdig} operate by executing containers and profiling the system events\footnote{Throughout this paper, we use \emph{system events} to refer to both \emph{syscalls} and \emph{capabilities}. When only syscalls are mentioned, it is to highlight the focus of prior work on this subset.} triggered during runtime. In contrast, static analysis-based approaches~\cite{ghavamnia2020confine,Chestnut,demarinis2020sysfilter} infer these events by analyzing the source code or binaries of containerized applications. However, both techniques suffer from fundamental limitations. Dynamic analysis fails to observe syscalls along execution paths that are not exercised during profiling, resulting in an \emph{underestimation} of the container’s requirements and potential execution failures. Conversely, static analysis disregards the actual container runtime behavior and conservatively includes all syscalls that might be invoked, leading to an \emph{overestimation} of required privileges and an unnecessarily large attack surface. Moreover, most existing solutions largely ignore Linux capabilities—another critical component of the container privilege model.

In this paper, we propose \ourtool{}, a novel attack surface reduction tool aimed at automating the identification of syscalls and capabilities invoked by containerized applications. The goal of \ourtool{} is to develop minimal yet effective security policies, surpassing the overestimation of static analysis-based solutions, while also addressing coverage issues inherent in existing dynamic analysis-based solutions. To achieve this goal, \ourtool{} relies on dynamic analysis, but distinguishes itself from previous solutions by considering the diverse execution environments to which containers may be exposed in real-world scenarios.

Specifically, \ourtool{} begins by generating a multitude of container execution environments, taking into account the combination of workloads subjected to containerized applications and their runtime configurations. Subsequently, it iteratively executes containers within the generated environments, capturing detailed system events to identify the necessary privileges during both container initialization and workload serving phases. This design enables \ourtool{} to profile diverse system events invoked by containers required for normal execution and corner cases, such as exception handling routines, thus overcoming the coverage limitations of existing dynamic analysis-based solutions.
Finally, the collected events are consolidated into final policies through our novel merging algorithm, which considers both the security and functionality guarantees of the generated policies. 

Profiling multiple execution environments might initially appear straightforward, leveraging prior dynamic analysis-based solutions. However, this approach presents significant challenges. First, container behavior is shaped by numerous environmental factors, including Docker command options and workloads, leading to an expansive profiling space akin to the state explosion problem in fuzzing. Second, while examining all possible values of Docker command options can reveal more system events, the process is time-consuming due to the extensive configuration and execution required for each container. To overcome these challenges, we propose two heuristics to efficiently navigate the environment space: (i) we demonstrate that a superset of system events can be inferred by analyzing its subsets independently, without requiring full execution; and (ii) we introduce an algorithm that efficiently explores the value range of a Docker option by dynamically adjusting the step size based on observed system events.

We developed a fully functional prototype of \ourtool{} using eBPF. To assess its effectiveness, we conducted experiments with official Docker container images, focusing on the additional syscalls identified when considering Docker command options and workloads. Result from the top 15 container images shows that \ourtool{} discovered 11\% and 22\% more syscalls when accounting for Docker command option and workload, respectively. Furthermore, we demonstrated its scalability, as all containers exhibit only a single syscall difference when merging different subsets. Lastly, through a security analysis involving 45 known vulnerabilities (e.g., CVEs), we demonstrated that \ourtool{} significantly reduces the attack surface, protecting an average of 87\% of containers from known vulnerabilities compared to 60\% with static analysis-based solutions.\\

\noindent\textbf{Contributions.} The main contributions of our work are:

\begin{itemize}
    \item We design and implement \ourtool{}\footnote{All our code is available at the following link: \url{https://github.com/haney1357/BeaCon.git}}, a practical tool that automatically generates container policies compatible with Linux Seccomp and Capabilities.

    \item We propose a novel approach for container policy generation based on emulated container environments to overcome the limitations of prior dynamic analysis techniques. We design efficient heuristics for environment exploration, an eBPF-based monitor to capture syscalls and capabilities, and security and functionality scores for intention-aware policy tuning.

    \item We evaluate \ourtool{} on real-world Docker containers, demonstrating its ability to generate effective, low-overhead policies and to mitigate realistic kernel-level attacks more successfully than existing solutions.
\end{itemize}

\section{Background and Motivation}
\label{ref_motivation}

In this section, we motivate our approach by explaining background on container hardening and why \ourtool{} is necessary for effective attack surface reduction in containers.\\

\noindent\textbf{Linux Container Hardening.} Limiting the privileges available to a container has proven to be an effective technique for minimizing the risk of kernel exploitation by malicious containers. 
Linux provides complementary mechanisms: seccomp to restrict which syscalls and their arguments can be made, capabilities to decompose the system root privilege into fine-grained privileges, and LSM/MAC, such as SELinux or AppArmor, for object-level policy. We scope \ourtool{} to seccomp and capabilities because they are widely supported by various container runtimes and orchestration tools (e.g., Kubernetes~\cite{kubernetes, kubernetes-seccomp}), map directly to runtime observable events (see \ref{subsec:event_monitoring}), and can be tightened without deployment-specific labeling or policy authoring required by LSM or MAC, which are complementary and left as orthogonal future work~\cite{SELinux,SELinux-RHEL, AppArmor,AppArmor-Ubuntu}. In cloud environments, Docker containers are typically launched with a default Seccomp policy, which disables only 44 syscalls out of more than 300+ allowed~\cite{docker-profile}, and a default set of capabilities (specifically 14)~\cite{docker-default-capability}. However, in both cases, users can define custom policies.\\

\noindent\textbf{Need for Restricting Both Syscalls and Capabilities.} While Seccomp is a powerful tool for restricting syscalls in containerized environments, it alone is insufficient to define fine-grained policies that fully protect the host OS kernel. For example, allowing the \texttt{socket} syscall—essential for many network applications—can expose the system to attacks if a compromised container creates raw sockets to target neighboring containers~\cite{nam2020bastion}. Simply disallowing the \texttt{socket} syscall could mitigate this risk, but would also break functionality for many legitimate containers. A more precise alternative is to revoke the \texttt{CAP\_NET\_RAW} capability, which is specifically required to create raw sockets. This example underscores the importance of considering both syscalls and capabilities for effective policy enforcement. However, most existing solutions\footnote{Refer to Section~\ref{sec:related_work} for a detailed comparison with existing work.}~\cite{lei2017speaker,Timeloops,podman,sysdig,ghavamnia2020confine,Chestnut,demarinis2020sysfilter,blair2023automated, zhan2022shrinking} focus solely on syscalls, highlighting the need for an approach that can automatically and accurately identify the full set of syscalls and capabilities required by a container.\\

\noindent\textbf{Need for Environment-Aware Container Profiling.}
Existing dynamic analysis-based solutions~\cite{lei2017speaker,Timeloops,podman,sysdig} suffer from limited coverage because they fail to capture system events triggered by \emph{environments}\footnote{In this paper, we define the term \textit{environment} as the set of external factors that influence container behavior and, consequently, the system events it invokes.} that are not exercised during the profiling phase. This limitation is particularly evident in applications with infrequently executed code paths (e.g., error-handling routines) or dynamic behavior (e.g., under specific workloads or high traffic). As a result, many such behaviors go undetected, creating blind spots in system event analysis. To demonstrate this, we analyze how environments elicit additional system events in ``redis'' and ``nginx'' containers. Table~\ref{tab:sytem_event_variation} summarizes the additional syscalls and capabilities observed under two Docker options and two workload profiles:

\begin{table}[t]
\small
\centering
\caption{System event variation in Nginx and Redis containers caused by different configuration/workload.}
\resizebox{\linewidth}{!}{
\begin{tabular}{c l c}
\toprule
    \textbf{\textbf{Image}} &\textbf{\textbf{Docker Options/Workload}} & \textbf{\textbf{Additional System Events}}\\ \midrule
    \texttt{redis} & -{}-init & \texttt{rt\_sigtimedwait}, \texttt{setpgid} \\
    \texttt{nginx} & -{}-network host & \texttt{NET\_BIND\_SERVICE} \\
    \texttt{redis} & Read \& Write & \texttt{fsync}, \texttt{fdatasync}, \texttt{fadvise64} \\ 
    \texttt{redis} & Bulk Read \& Write & \texttt{writev}, \texttt{shutdown}, \texttt{sync\_file\_range} \\
\bottomrule
\end{tabular}
}
\label{tab:sytem_event_variation}
\end{table}

\begin{itemize}
    \item \textbf{Init process (-{}-init): } Many orchestrations inject a tiny init (e.g., \texttt{tini}) to reap zombies and forward signals. It creates a new process group (\texttt{setpgid}) and waits for signals (\texttt{rt\_sigtimedwait}). If a seccomp profile without these syscalls, the container fails at startup with an error message: Operation not permitted.
    \item \textbf{Binding privileged ports (-{}-network host): } Binding to ports less than 1024 is privileged and requires \texttt{CAP\_NET\_BIND\_SERVICE}. As Nginx is built to expose port 80 in default, letting container to use the host network results binding to port 80. The consequences of dropping \texttt{CAP\_NET\_BIND\_SERVICE} cause a permission denied error in \texttt{bind} syscall.
    \item \textbf{Ordinary read \& write workload: } In default, Redis turns on persistence, which activates the file-persistence path such as sync to disk, advise access pattern. When Redis receives workload, it begins to synchronize data to the disk, which fail with the message, "Operation not permitted", if the syscalls such as \texttt{fsync}, \texttt{fdatasync}, \texttt{fadvise64} are blocked.
    \item \textbf{Bulk/pipelined writes: } Large and pipelined writes trigger vectored I/O (\texttt{writev}) and explicit socket teardown (\texttt{shutdown}). If these are omitted, bulk insert degrades or fails.
\end{itemize}

These system event differences are not rare corner cases; they are driven by common deployment choices (init, port) and load shapes. A policy that is correct in a single configuration can crash or silently degrade in another. This motivates our design of sweeping representative options and workloads (Section~\ref{subsec:emulating}) to explicitly emulate environments.\\

\noindent\textbf{Need for Balancing Security and Functionality.}
As shown in the arangodb container example, selecting which syscalls to include in security policies is straightforward in some cases: syscalls that are never invoked can be safely excluded, while those consistently used are necessary for maintaining functionality. The challenge arises with syscalls that are invoked only sporadically in specific environments, where the decision to include or exclude them is neither trivial nor universally applicable. Including such infrequent syscalls may unnecessarily expand the attack surface, as noted in static analysis-based solutions~\cite{ghavamnia2020confine}, whereas excluding them can lead to container failures or degraded functionality, a known issue in dynamic analysis-based approaches that capture only a limited set of system events~\cite{lei2017speaker}. Addressing this trade-off requires application-specific decisions that carefully weigh security risks against functional requirements for each system event.

\section{Threat Model and Assumptions}
\label{sec:threat_model}

\noindent\textbf{Threat Model.} We consider adversaries who have gained access to a container, typically by exploiting a vulnerability in the container application, although the specific method is beyond the scope of this document. Studies~\cite{10.1145/3029806.3029832,10.1007/978-3-030-58951-6_13} have shown that many container images in public repositories, such as Docker Hub, including popular ones downloaded millions of times, contain critical vulnerabilities that often remain unpatched for extended periods. Once inside, the goal of adversaries is to escape the container's isolation to compromise co-resident containers or gain elevated privileges on the host system, similar to the lateral movement or malware propagation~\cite{yu2014malware,matta2018cyber,karyotis2015macroscopic}. Achieving this typically involves exploiting a kernel vulnerability~\cite{10.1145/3274694.3274720}, for which the adversary can abuse any privileges granted to the container. Thus, the broader the container-kernel interface (i.e., the more syscalls allowed), the larger the attack surface, increasing the likelihood of finding and exploiting a kernel vulnerability.\\

\noindent\textbf{Assumptions.} We assume that containers are executed without root privileges and are properly isolated using standard OS-level security mechanisms provided by the Linux kernel (e.g., cgroups~\cite{cgroups} and namespaces~\cite{namespace}), which are common practices in containerized environments. Our primary assumption is that container images are not compromised when pulled from image repositories. This is reasonable, as cloud providers typically employ security scanning tools to detect vulnerabilities before deployment~\cite{securitytools}. This assumption ensures that containers remain uncompromised during our offline profiling phase (Section~\ref{sec:beacon}), and thus, no attacks are performed during this stage. We further assume that reducing the attack surface of operating system (e.g., Ubuntu) and language runtime (e.g., Python) container images is not feasible. These base images are not intended to directly support specific service workloads but rather serve as foundations for building application-specific containers. In practice, tenants typically deploy customized images derived from these bases, rather than using them directly. Therefore, we exclude such base images from the set of target containers in our study.

\section{BeaCon Design}
\label{sec:beacon}
\begin{figure}[htbp]
    \centering
    \centerline{\includegraphics[width=\linewidth]{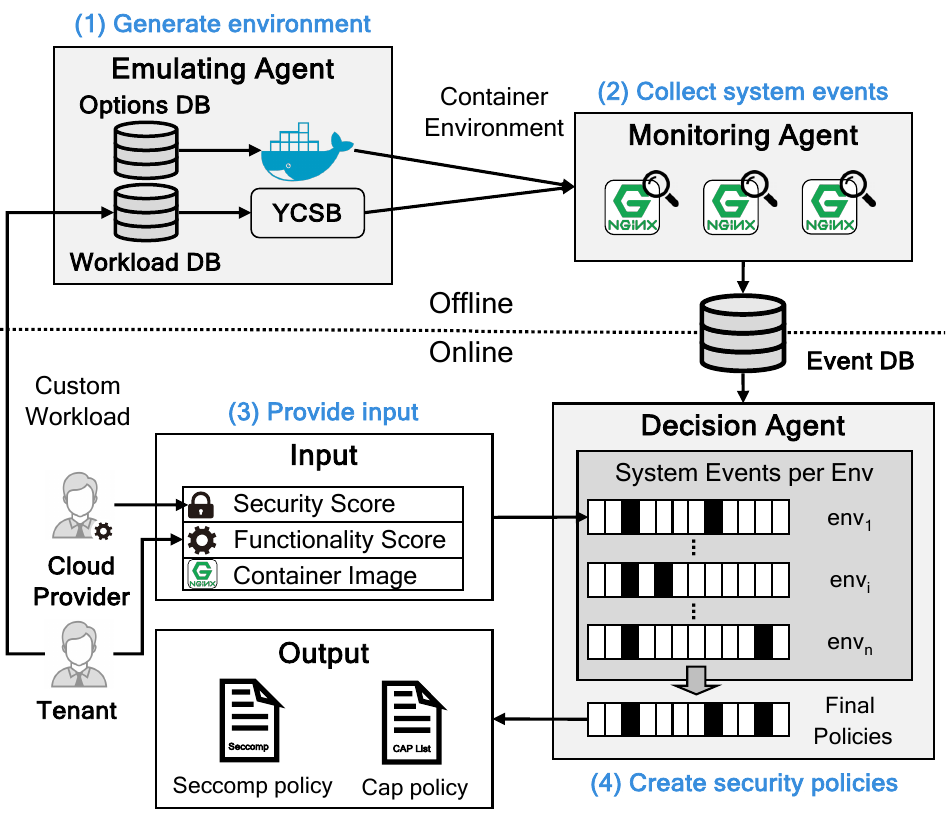}
}
    \caption{Architecture of \ourtool{}. }
    \label{f:my_label}
\end{figure}

The goal of \ourtool{} is to enable cloud providers to automatically identify which system calls and capabilities should be allowed or blocked during container execution, ensuring security without compromising functionality.\\

\noindent\textbf{Design Rationale.} \ourtool{} is designed based on three principles: portability, observability, and low operator friction. First, we target seccomp-BPF and Linux capabilities because these controls are universally supported across OCI runtimes/Kubernetes and map directly to runtime-observable signals (syscalls at sys\_enter, capability checks via cap\_capable). This enables quantifying coverage and tune policies without deployment-specific labeling or device rules. Second, we introduce environment emulation as an explicit step: \ourtool{} systematically varies common container deployment options and workloads, so that environment-conditioned events are triggered and captured. Third, we formulate policy synthesis as a dual-criteria selection (functionality vs. security) with operator-tunable scores, yielding a frontier of policies rather than a single fixed profile. Compared to prior static allow-list generators, \ourtool{}’s contribution is the environment-aware discovery and tunable synthesis pipeline that (i) recovers rarely triggered but necessary events, (ii) exposes transparent trade-offs, and (iii) requires no application code access or MAC/device policies.\\

\noindent\textbf{Overview.}
\ourtool{} consists of three main components: (i)~the \emph{Emulation Agent}, (ii)~the \emph{Monitoring Agent}, and (iii)~the \emph{Decision Agent} (see Figure~\ref{f:my_label}). The overall procedure unfolds as follows. As a preliminary step, the Emulation Agent builds an Option and Workload database, which is used to emulate diverse container environments.
(1)~Leveraging the container engine, the Emulation Agent executes the target container image using configuration options stored in the Option DB. It then applies workloads from the Workload DB using YCSB, a benchmarking tool equipped with various client types for generating API requests. This process is repeated for each combination of options and workloads.
(2)~The Monitoring Agent observes the system events generated by each emulated container using eBPF and stores the collected data in the Event DB.
(3)~Next, the Decision Agent gathers security and functionality scores provided by both the cloud provider and tenants.
(4)~Using the system events stored in the Event DB along with the provided scores, the Decision Agent generates security policies that (i)~allow only the system events consistently observed across environments and (ii)~satisfy the specified security and functionality constraints.\\

\noindent\textbf{Online and Offline Phase.}
At first glance, \ourtool{} could be designed to operate entirely online, profiling containers immediately upon receiving a request to start them. However, the profiling phase in \ourtool{} requires running containers multiple times—each under a different environment—and collecting system events over several minutes per run. For instance, if a cloud provider considers 10 different environments and monitors each for 5 minutes, the total profiling time quickly adds up. In practice, the monitoring duration may be longer, and many containers may need to be profiled simultaneously. This introduces significant latency, which can hinder the fast startup times expected in containerized environments. To address this, \ourtool{} adopts a two-phase approach: an \emph{offline} phase and an \emph{online} phase. In the offline phase, \ourtool{} executes containers across both generated and user-defined environments, collecting and storing the resulting system event data. This pre-collected information enables fast, informed policy decisions in the online phase, where container profiles are applied in real-world deployment scenarios with minimal delay.
The remainder of this section introduces the design details of \ourtool{}.

\subsection{Container Environment Emulation}
\label{subsec:emulating}

The Emulation Agent is responsible for generating a set of environments that containers are likely to encounter and executing the containerized application under each environment, one at a time. Ideally, constructing such environments would benefit from a deep understanding of the application's internal behavior, including how it is launched and how end users interact with it. However, this is impractical in real-world settings due to the vast number and diversity of containerized applications, the complexity of their implementations, and the fact that their source code is often unavailable to cloud providers.

To overcome this limitation, \ourtool{} adopts a black-box approach that does not rely on prior knowledge of application internals. It emulates execution environments by replicating key external factors that influence container behavior. In particular, it focuses on two critical dimensions: (i) Docker command options specified at container initialization, and (ii) workloads applied during runtime. This focus is motivated by the observation that these two factors have a significant impact on how containers behave in practice. As illustrated in the upper left of Figure~\ref{f:my_label}, \ourtool{} maintains two dedicated databases to support environment emulation—one for storing Docker command options supported by each container, and another for capturing representative workloads.\\

\begin{table}[t]
\small
\centering
\caption{Formal syntax of Docker command options as defined by \ourtool{}.}
\resizebox{\linewidth}{!}{
\begin{tabular}{c c l}
\toprule
\textbf{\textbf{Name}} & \textbf{\textbf{Type}} & \textbf{\textbf{Formal Syntax}}\\ \midrule
attach          & list      & \texttt{<List>:( ``stdin'' | ``stdout'' | ``stderr'')} \\
detach          & bool      & \texttt{<Bool>} \\
tty             & bool      & \texttt{<Bool>} \\
interactive     & bool      & \texttt{<Bool>} \\
stop-signal     & string    & \texttt{<Signals>} \\
health-retries  & int       & \texttt{<U32>} \\
publish         & list      & \texttt{<List>:} \\
                &           & \texttt{(<Continuous\_range>:(0, U16) ``:''} \\
                &           & \texttt{<Continuous\_range>:(0, U16) } \\
                &           & \texttt{[``/''(``tcp'' | ``udp'')])} \\
pids-limit      & int       & \texttt{``-1'' | <U22> } \\

memory          & bytes     & \texttt{<U32> [(``n'' | ``k'' | ``m'' | ``g'')] } \\
memory-swap     & bytes     & \texttt{<U32>} \\
memory-reservation    & bytes     & \texttt{<U32>} \\
kernel-memory   & bytes     & \texttt{<U32>} \\
cpu-shares      & int       & \texttt{<U18>} \\
cpu-period      & int       & \texttt{(1000, 1000000)} \\
stop-timeout    & int       & \texttt{<U32>} \\
volume          & list      & \texttt{<List>:([<HVPath>``:'']} \\
                &           & \texttt{<CVPath>[``:''[``ro'' | ``rw'']])} \\
oom-score-adj   & int       & \texttt{<I11>} \\
shm-size        & bytes     & \texttt{<U32>} \\

\bottomrule

\end{tabular}
}
\label{tab:UsedOption}
\end{table}

\noindent\textbf{Container Runtime Options Database.}
\ourtool{} constructs and maintains a database of valid Docker command options for each container, along with their observed impact on the system events generated during execution. To build this database, we first gathered information about existing Docker options, including their types and the range of permissible values. This step is critical, as omitting it would lead the Emulation Agent to generate invalid Docker commands. We excluded five options related to security enforcement or informational help (e.g., \texttt{cap-add}, \texttt{cap-drop}, and \texttt{help}) as they are either outside the scope of emulation or not meaningful for behavior profiling. While creating this database, we observed that the current classification of Docker option types~\cite{docker-default-capability} is not sufficiently expressive. This limitation stems from the fact that the official documentation is not written in a fully formal or structured format, making it difficult to accurately infer valid values and types for each option.\\

\noindent\textbf{Defining Formal Syntax for Container Options.}
To address the limitations of informal documentation, we developed a formal syntax to define container options, explicitly specifying their types and permissible value ranges. For example, we introduced a new type, \texttt{<U18>}, for the \texttt{cpu-shares} option, which was previously labeled merely as an “integer” with a vaguely defined lower bound of 2~\cite{docker-default-capability}. To determine the upper bound, we incrementally tested powers of two, starting from $2^1$ up to $2^{32}$, stopping when the container failed to launch. The container failed at $2^{18}$, indicating that the value must be an unsigned integer less than $2^{18}$. Another example involves options that accept compound types, such as \texttt{memory}, which is documented as \texttt{bytes} but accepts a numeric value optionally followed by a unit suffix (e.g., bytes, kilobytes, megabytes, or gigabytes). We formally defined it as \texttt{<U32>["b"|"k"|"m"|"g"]}, meaning it accepts any positive integer less than $2^{32}$ followed by one of the specified unit characters. This formal representation ensures that each option’s value space is finite, enabling \ourtool{} to randomly generate and mutate values while maintaining syntactic validity and ensuring comprehensive coverage. A summary of the formal syntax is presented in Table~\ref{tab:UsedOption}, and additional details are provided in our supplementary material~\cite{ourfolder}.\\

\noindent\textbf{Workloads Database.}
In addition to maintaining a database of valid Docker options and their effects on container behavior, \ourtool{} also constructs and maintains a database of representative workloads. It distinguishes between two main types of container workloads: (i) those with common and standardized APIs (e.g., typical operations in a database), and (ii) those with custom APIs (e.g., HTTP RESTful APIs exposed to end users). In both cases, there are multiple ways for the Emulation Agent to identify relevant workloads to test. One approach is for tenants to provide the workloads directly—especially valuable when dealing with containers exposing custom APIs. Alternatively, cloud providers can reuse real-world workloads observed from pre-released containers. Given that providers already operate large numbers of containers in production, they can intercept traffic between containers and clients, extract the exchanged message streams, and reuse these as representative workloads. Finally, cloud providers can leverage existing benchmarking tools to artificially generate workloads. Although these tools are primarily designed to evaluate container performance, they are highly effective at triggering diverse execution paths and behaviors that might otherwise be difficult to observe.

\subsection{Environment Emulation Heuristics}
\label{subsec:heuristics}

Ideally, \ourtool{} would emulate all possible combinations of container environments to ensure comprehensive coverage. However, this approach incurs substantial computational and storage overhead, making it impractical at scale. To address this, we design heuristics that reduce the profiling space by selectively limiting the number of environment combinations in which each container must be executed, while still preserving profiling effectiveness.\\

\begin{algorithm}[!t]
    \footnotesize
    \caption{Option value mutation}\label{alg:randmutate}

    \SetKwInput{KwInput}{Input}
    \SetKwInput{KwOutput}{Output}
    \SetKwFunction{monitor}{monitor}
    \SetKwFunction{random}{random}
    \SetKwFunction{normal}{normal}

    \KwInput{ 
        The target environment $env$, \\
        The minimum option value \( v_{min} \),\\ 
        The maximum option value \( v_{max} \),\\
        A scaling factor for adjusting the step size \( r \),\\
        The initial step size \( step_{init} \),\\  
        The maximum number of iterations 
        \( it_{max} \),\\
        The probability of resetting the option value \( p \)
        }

    \KwOutput{
        A set of recorded system events $E$   
    }
    
        $v \gets \random{$v_{min}, v_{max}$}$ 
        
        $step \gets step_{init}$
    
        $t_{base\_lower} \gets 5$
        
        $t_{base\_upper} \gets 10$
        
        $\lambda \gets 0.03$
         
        $E \gets \emptyset$
        
        $it \gets 0$

    \For{$v \leq v_{max}$ and $it < it_{max}$}{
        $E' \gets \monitor{$env, v$}$

        $t_{lower} \gets t_{base\_lower} \times e^{-\lambda \times it}$
        
        $t_{upper} \gets t_{base\_upper} \times e^{-\lambda \times it}$
        
        \If {$|E' - E| < t_{lower}$}{
            $step \gets step \times r \times (1 + \normal(\mu, \sigma))$
        }
        \ElseIf {$|E' - E| \geq t_{upper}$}{
            $step \gets step / r$
        }

        $v \gets v + step$
        
        \If {$\random(0, 1) < p$}{
            $v \gets \random(v_{min}, v_{max})$
        }

        $E \gets E \cup E'$
        
        $it \gets it + 1$
    }
\end{algorithm}

\noindent\textbf{Randomized Testing Environment Mutation.}
The Emulating Agent selects values for options or workloads to emulate a container environment, repeatedly generating environments with varying option values or workloads to observe the container's behavior under diverse configurations. However, the vast range of possible values for some options, particularly integer types like \texttt{<U32>} with billions of potential values, makes exhaustive testing impractical. To address this, we propose a random mutation algorithm, detailed in Algorithm~\ref{alg:randmutate}, which heuristically explores option values that may trigger system events unique to containers, inspired by fuzzing techniques. The algorithm's goal is to efficiently navigate the environment space, enabling the discovery of additional system events while strategically mutating option values. To achieve this, it dynamically adjusts the step size for mutating an option value based on the number of new system events observed.

The algorithm begins by randomly selecting an initial option value $v$ from the range between the minimum $v_{min}$ and the maximum $v_{max}$ (line 1). These bounds, $v_{min}$ and $v_{max}$, are derived from Table~\ref{tab:UsedOption} which documents the range of valid values for each option. It then repeatedly mutates the value $v$ to explore new configurations and monitors a new system event set $E'$ (lines 8-9). At each iteration, the algorithm calculates the upper threshold $t_{upper}$ and lower threshold $t_{lower}$, with the values decaying exponentially as the iteration $it$ progresses (lines 9-11). These thresholds determine the next step size based on the number of new system events, denoted as $|E' - E|$. If the number is below the lower threshold, the step size is increased using a scaling factor and a Gaussian perturbation to prioritize exploring further values (lines 12-13). Conversely, if the difference in system events exceeds the upper threshold, the step size is reduced, as the current step size is likely too wide to identify meaningful values (lines 14-15). The algorithm mutates the option value based on the current value and step size (line 16). 

To avoid biased exploration caused by initial option value, the algorithm randomly resets the value with a predefined probability $p$ (lines 17-18). This random reset encourages broader exploration of the option space. The random mutation continues until the option value $v$ exceeds the valid range or the iteration count reaches predefined limit (line 8). By employing this algorithm, the Emulating Agent efficiently explores a wide range of integer-type option values while limiting the number of tests to a practical size.  This approach ensures the discovery of meaningful configurations that can invoke system events not observed in other environments, enabling comprehensive testing of container behavior.\\

\noindent\textbf{Optimization for Reducing Emulation Space.}
Despite the use of our algorithm described above, the vast number of possible environment combinations (e.g., Docker command options and workloads) presents a significant scalability challenge for \ourtool{}. Exhaustively emulating all combinations is computationally infeasible due to the large emulation space and time requirements. To address this, we propose an optimization strategy based on the following rationale: suppose the system event sets $E_i$ and $E_j$ have been profiled for environments $i$ and $j$, respectively. Our goal is to estimate the system event set $E_k$ for a new environment $k$. If environment $k$ combines attributes from both $i$ and $j$ (where $i \neq j$), we infer that $E_k = E_i \cup E_j$, eliminating the need to emulate $k$ directly. This inference-based approach enables \ourtool{} to efficiently generate comprehensive security policies without executing or storing data for every possible container-environment combination. The feasibility of this method is demonstrated in Section~\ref{sec:ev_indp}.

\subsection{Container Event Monitoring}
\label{subsec:event_monitoring}

The Monitoring Agent executes the container multiple times, applying each environment generated by the Emulation Agent one at a time. For each environment, it collects the system calls invoked and the capabilities requested by the container. To achieve this, we leverage extended Berkeley Packet Filter (eBPF)~\cite{ebpf}, which enables the Monitoring Agent to be triggered by predefined events such as system calls or network activity. These events are captured using Linux kernel probes like Kprobes and Tracepoints. \ourtool{} utilizes these eBPF features to extract detailed information about the syscalls and capabilities requested during container execution.\\

\begin{figure}
    \centering
    \includegraphics[width=.9\linewidth]{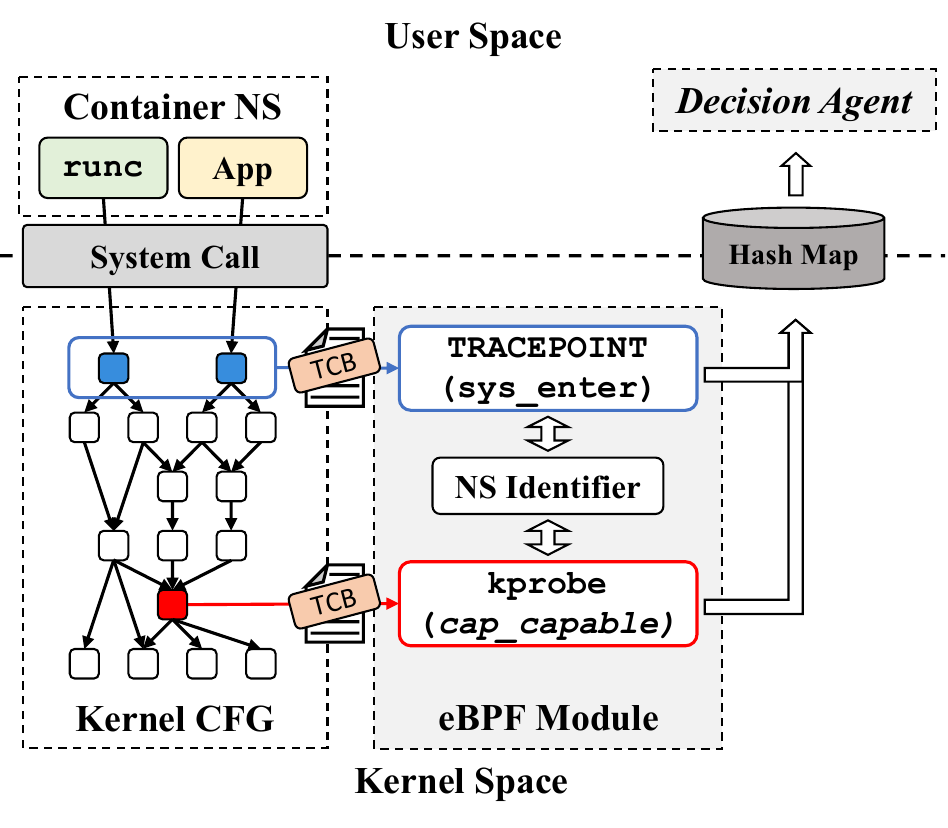}
    \caption{Container event monitoring architecture in \ourtool{}. System calls are intercepted using \texttt{sys\_enter} with Tracepoints, and capabilities are captured using \texttt{cap\_capable} with Kprobes.}
    \label{fig:monitoring}
\end{figure}

\noindent\textbf{Configuring Tracing.} To begin with, we define four Tracepoint probes to monitor the occurrences of four syscalls, namely \texttt{unshare}, \texttt{seccomp}, \texttt{prctl}, \texttt{capset}. The reason why we attach a probe to the \texttt{unshare} syscall is because it is invoked when a namespace is created. Since each time a new container is executed a namespace is created that is associated with it, we use this syscall as a reference point to detect the launch of a new container. The remaining probes are used to detect the exact moment when the container becomes controlled by the Seccomp and capabilities Linux mechanisms. Before this, containers can generate certain system events but these should not be considered in the security policy generation. In addition to the previous, we defined two additional probes, \texttt{sys\_enter} and Kprobe (attached to \texttt{cap\_capable}), respectively, in order to monitor all syscalls and capabilities requested by the containerized application. Finally, due to the volatility of local variables in probe controller logic that cannot be accessed from other probes, we created an eBPF hash map where keys are namespaces representing only containers and values. The values consist of two flags, \texttt{seccomp\_flag} and \texttt{capability\_flag}, indicating the state of the container, and two bitmaps for storing the requested capabilities and syscalls.\\

\noindent\textbf{Monitoring System Events.} Once the eBPF program starts running, the defined probes capture all system events, waiting for an incoming \texttt{unshare} syscall. Upon this syscall, the eBPF program extracts the namespaces from the originating process, then creates a new entry in the eBPF hash map with a namespace key. The flags \texttt{seccomp\_flag} and \texttt{capability\_flag} are set to ``False'', and the bitmaps are set to 0. The eBPF program then waits for \texttt{prctl} or \texttt{seccomp} syscalls and changes the state of the flags from False to True when these syscalls are received. From this point onward, the eBPF program begins marking the syscalls and capabilities perceived by \texttt{sys\_enter} of Tracepoints and \texttt{cap\_capable} of Kprobes into the hash map, respectively, as illustrated in Figure~\ref{fig:monitoring}. The Monitoring Agent ignores all syscalls that originate from namespaces that do not match the namespace of the container being monitored.

\subsection{Container Policy Decision}
\label{sec:decision_agent}

The Decision Agent determines an appropriate policy based on the collected system events, while balancing the container's security and functionality. As discussed, this process is non-trivial and must account for the specific risk and operational requirements of cloud providers or tenants.\\

\noindent\textbf{Security and Functionality Scores.}
To support this trade-off, we introduce a quantitative framework consisting of a \emph{security score} and a \emph{functionality score}. These scores enable cloud providers or tenants to customize policies that align with their security and availability requirements.

The security score \( S^{\text{sec}}_{c} \) for a container \( c \) is defined as:
\begin{equation}
    S^{\text{sec}}_{c}(E) = 1 - \max_{e \in E}{ \frac {CVSS(e)}{10}}
    \label{eq:security_score}
\end{equation}
where \( CVSS(e) \) denotes the highest CVSS (Common Vulnerability Scoring System) value among the Linux kernel CVEs (Common Vulnerabilities and Exposures)~\cite{CVE} associated with system event \( e \in E \), and \( E \) is the set of system events (i.e., system calls or capabilities) granted in the policy. The score is normalized by dividing the CVSS value by 10, reflecting its original range of 0 to 10. This formulation captures the worst-case severity among the allowed events, serving to highlight risky events—not to blindly block them. 

The use of CVSS offers several benefits: it provides access to a comprehensive history of known vulnerabilities through CVEs, employs a standardized language for linking vulnerabilities to specific system events, and uses a well-established framework for quantifying risk. Following the methodology of Confine~\cite{ghavamnia2020confine}, the Decision Agent maps Linux kernel CVEs to system events observed in each container. Nonetheless, this method cannot account for zero-day exploits, which lack CVSS scores and therefore remain unassessed (see Section~\ref{sec:discussion}). Our goal, however, is to proactively reduce the container's attack surface by leveraging historical vulnerability data to assess and prioritize risk.

The functionality score \( S^{\text{func}}_{c} \) is defined as:
\begin{equation}
    S^{\text{func}}_{c}(E) = \frac {|\{x \in X \mid E^{'}_{c}(x) \subseteq E\}|}{|X|}
    \label{eq:functionality_score}
\end{equation}
where \( E'_{c}(x) \) denotes the set of system events triggered when the external environment \( x \in X \) is applied to container \( c \). This score measures the proportion of environments in which the container can operate normally under the given policy \( E \), serving as an approximation of runtime coverage. To compute this score, the Decision Agent uses data collected by the Monitoring Agent, which captures the frequency of system events observed during container execution across diverse environments. This enables the system to identify rarely used but essential events and avoid overly restrictive policies that might compromise container functionality.\\

\noindent\textbf{Policy Generation.}
The security score ensures that the policy excludes system events associated with CVEs whose CVSS scores exceed the threshold specified by the cloud provider. To reflect this in policy generation, the Decision Agent examines all system events observed by the Monitoring Agent, considering their associated vulnerabilities, exploitability, and CVSS scores. It then includes only those events whose CVSS values are below the specified threshold, thereby ensuring that the container is not exposed to vulnerabilities deemed more severe than the acceptable risk level.

The security and functionality scores act as guiding metrics for determining which system events to include in the final policy. As illustrated in the lower right portion of Figure~\ref{f:my_label}, the Decision Agent generates the policy based on system events collected from each environment. Events that are never observed in any environment are excluded by default, while those consistently observed are unconditionally included. For events that appear sporadically, the Decision Agent evaluates their inclusion by checking whether the resulting policy satisfies both the security and functionality thresholds. If no such policy can be generated under the given constraints, \ourtool{} reports the failure and prompts the user to revise the thresholds.

\section{Evaluation}
\label{sec:evaluation}

In this section, we evaluate the effectiveness of \ourtool{} in generating policies that are accurate, secure, and lightweight. Our evaluation is guided by the following key questions:
\begin{itemize}
\item \textbf{Q1.} Can the environments emulated by \ourtool{} help uncover additional system events? (Section~\ref{sec:ev_diff_opt})
\item \textbf{Q2.} Do \ourtool{}'s emulation heuristics reduce the profiling space? (Section~\ref{sec:ev_indp})
\item \textbf{Q3.} Can functionality and security scores assist in policy fine-tuning? (Section~\ref{sec:ev_need_of_spec})
\item \textbf{Q4.} What is the performance overhead of \ourtool{}'s event monitoring in containers? (Section~\ref{sec:ev_perf})
\item \textbf{Q5.} How does \ourtool{} compare to static analysis-based solutions in reducing attack surfaces? (Section~\ref{sec:ev_CVE})
\item \textbf{Q6.} Is \ourtool{} effective in protecting containers against known attacks? (Sections~\ref{sec:known_attacks} and~\ref{sec:securityusecases})
\end{itemize}

\subsection{Experimental Setup}
\label{sec:implementation}

\noindent\textbf{Implementation.} We implemented a full-fledged prototype of \ourtool{} with 1320 lines of Python code for the Emulation and Decision Agent. For the Monitoring Agent, we employed the BPF Compiler Collection (BCC) tool-chain, a Python framework that simplifies writing eBPF programs~\cite{bcc}. We developed eBPF programs for the Monitoring Agent with 170 lines of C code. All our experiments were conducted on a server with a 4-core Intel i5-6600K 3.50 GHz CPU and 16GB of RAM. We used Ubuntu 20.04.3 LTS with the Linux kernel v5.4.0-89 and the Docker Engine v20.10.11.\\

\noindent\textbf{Dataset.} We collected 178 Docker container images from Docker Hub~\cite{docker_hub}, a publicly available repository that hosts container images from various providers. To ensure the successful execution of containers, we developed a crawler specifically design to collect ``official'' images maintained by Docker. We then filtered out unsupported containers by examining their tag (i.e., container image version) and selected those that were compatible with our testbed running the Linux AMD64 architecture. From the collected images, 11 containers were not compatible with our environment, so we ended up using 167 container images in our experiments (unless specified otherwise in a given experiment).\\

\noindent\textbf{Workload Generation.} We utilized the widely recognized YCSB benchmark framework~\cite{YCSB} to evaluate container behavior under scenarios that are challenging to replicate outside real-world environments. YCSB provides multiple client instances tailored for database-type containers\footnote{Currently supported database containers include: Aerospike, ArangoDB, Cassandra, Couchbase, Elasticsearch, Memcached, MongoDB, OrientDB, PostgreSQL, Redis, and ZooKeeper.} and a generic REST-API client instance for web server-type containers. In addition to the client instances offered by YCSB, we developed custom database client instances for unsupported official database-type containers. YCSB supports common database operations such as read, update, scan, insert, and delete, and facilitates workload generation for web server containers using REST-API methods like GET, POST, and PUT. To evaluate the impact of workloads on container profiling, we defined specific workloads and configured their properties in YCSB. Table~\ref{tab:Workload_Prop} provides a detailed description of the properties used in the eight workloads ($W_1$–$W_8$), which were designed based on three key properties commonly encountered during the container lifecycle:
\begin{itemize}
    \item \textbf{Operation Types:} We included all available operations supported by YCSB, namely, Read, Update, Scan, Insert, and Delete. Each operation request consisted of 1,000 operations, which is approximately the maximum allowed by YCSB. This high number of operations aimed to exert extreme load on the containerized applications.
    \item \textbf{Query Record Properties:} The workloads encompassed varying types of data in client requests. For this purpose, we set different values for field count and field length. The default value of Field Count and Field Length were 10 and 100 respectively. Additionally, we used larger values of 500 and 10,000 for field count and length which are also close to the maximum supported by YCSB.
    \item \textbf{Thread Count}: The final workload was designed to simulate a high load scenario where a large number of clients simultaneously accessed the container. To achieve this, we utilized 500 threads to send requests concurrently.
\end{itemize}

\begin{table}[t]
\small
\centering
\caption{YCSB properties used to generate 8 different workloads. The table includes the types of operations performed by each workload, query properties (specifically relevant to RDBs), and the thread count, which enables concurrent requests from multiple threads.}
\resizebox{\linewidth}{!}{
\begin{tabular}{c c c c c c c c c}
\toprule

\multirow{2}{*}{\textbf{Index}} & \multicolumn{5}{c}{\textbf{{Operation Count}}} & \multicolumn{2}{c}{\textbf{Query Properties}} & \multirow{2}{*}{\makecell{\textbf{Thread}\\\textbf{Count}}} \\ 
\cmidrule(l){2-6} \cmidrule(l){7-8}

& Read & Update & Scan & Insert & Delete & Count & Length & \\ \midrule
$W_1$ & 1K & 0 & 0 & 1K & 0 & 10 & 100 & 1 \\
$W_2$ & 0 & 1K & 0 & 1K & 0 & 10 & 100 & 1 \\
$W_3$ & 0 & 0 & 1K & 1K & 0 & 10 & 100 & 1 \\
$W_4$ & 0 & 0 & 0 & 1K & 0 & 10 & 100 & 1 \\
$W_5$ & 0 & 0 & 0 & 1K & 1K & 10 & 100 & 1 \\
$W_6$ & 1K & 0 & 0 & 1K & 0 & 500 & 100 & 1 \\
$W_7$ & 1K & 0 & 0 & 1K & 0 & 10 & 10K & 1 \\
$W_8$ & 1K & 0 & 0 & 1K & 0 & 10 & 100 & 500 \\

\bottomrule
\end{tabular}
}
\label{tab:Workload_Prop}
\end{table}

\noindent\textbf{Scope and Generality.} We use YCSB to drive server-class images (databases, web, content-management-system (CMS), API management) because it provides a repeatable harness for automation. Note that this reflects evaluation practicality, not a methodological constraint. \ourtool{} is container-type agnostic; it only requires observable execution such as HTTP requests, CLI scripts, or replayed traces, after which the same eBPF monitor and policy generator apply unchanged.\\

\noindent\textbf{Applicability beyond DB/Web}. To justify the use of YCSB, we surveyed 167 container images across 13 categories and found that 94 images expose network ports by default (See Table~\ref{tab:container_category}). Of those port-exposed images, about 69\% (65/94) fall into server-style categories (database-and-storage, web-servers, content-management-systems, API-management, message-queues, monitoring-and-observability, and networking). Databases and web servers account for 55\% (36/65) of server-style categories. For non-DB/non-web containers, \ourtool{} uses the same interface but swaps the workload source: (i) an HTTP/REST for API services and CMS, (ii) scripted commands for batch/CLI tools (running entrypoint commands or test suites); and (iii) trace-ingest mode that profiles a container while replaying tenant-provided traces or production traffic. In all cases, \ourtool{}'s environment emulation (ports, init, volumes, read-only filesystems, etc.) remains applicable, and policy generation is unchanged.

\begin{table}[t]
\small
\centering
\caption{Survey of 167 container images across 13 categories (left: count per category; right: images that expose at least one port by default.}
\resizebox{\linewidth}{!}{
\begin{tabular}{l | c c}
\toprule
    \textbf{\textbf{Category}} &\textbf{\textbf{Number of images}} & \textbf{\textbf{Number of exposed images}}\\ \midrule
    data-science                    & 5     & 2     \\
    security                        & 2     & 1     \\
    \rowcolor{blue!20}
    api-management                  & 3     & 3     \\
    \rowcolor{blue!20}
    message-queues                  & 5     & 5     \\
    \rowcolor{blue!20}
    databases-and-storage           & 25    & 25    \\
    \rowcolor{blue!20}
    web-servers                     & 12    & 11    \\
    developer-tools                 & 8     & 3     \\
    networking                      & 1     & 1     \\
    languages-and-frameworks        & 34    & 4     \\
    monitoring-and-observability    & 7     & 7     \\
    \rowcolor{blue!20}
    content-management-system       & 14    & 14    \\
    operating-systems               & 17    & 0     \\
    integration-and-delivery        & 1     & 1     \\
    Uncategorized                   & 33    & 17    \\\hline
    Sum                             & 167   & 94    \\
\bottomrule
\end{tabular}
}
\label{tab:container_category}
\end{table}

\begin{figure}[t]
    \centering
    \includegraphics[width=\linewidth]{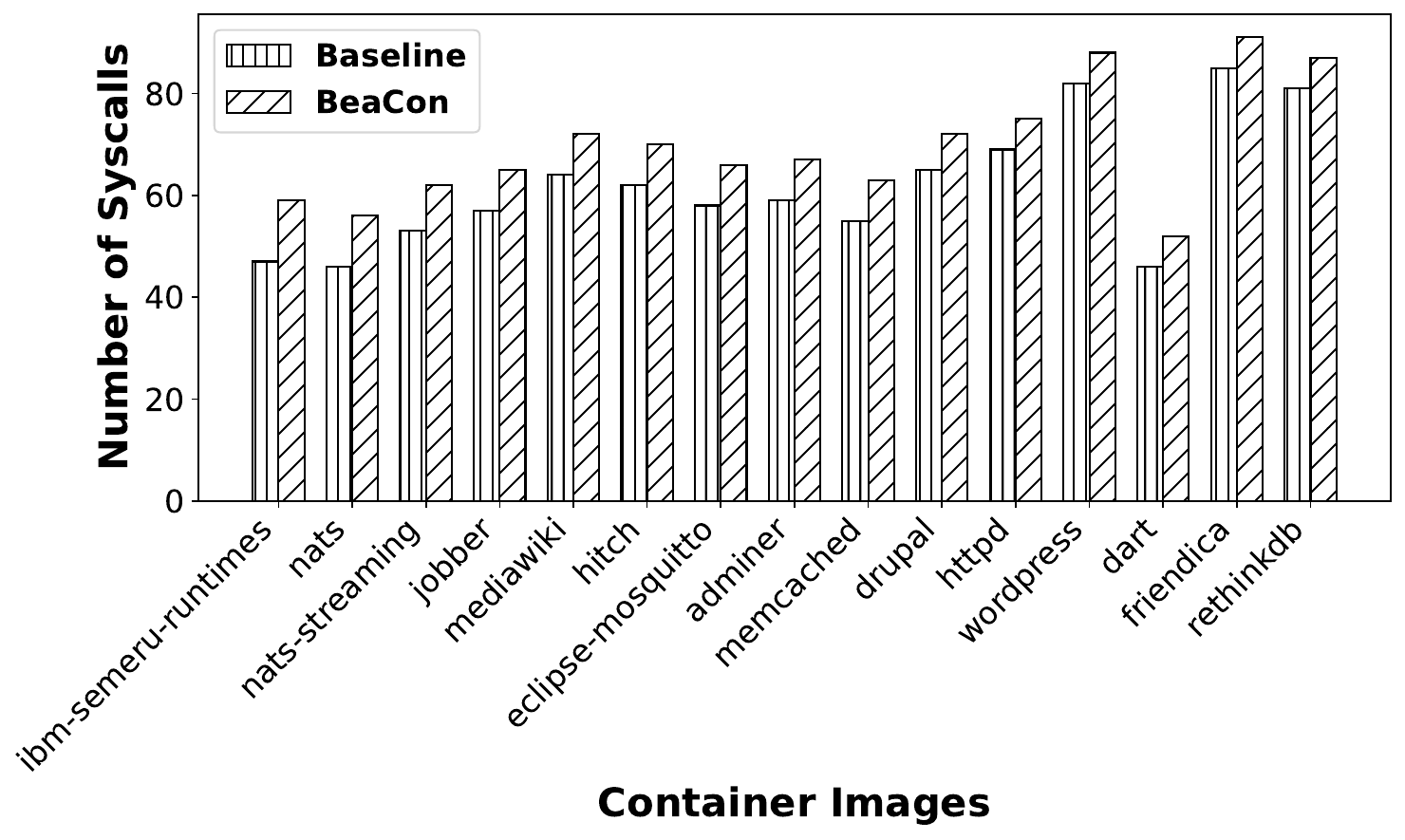}
    \caption{The number of syscalls observed without Docker options applied (\emph{baseline}) compared to those recorded when all tested options are enabled using \ourtool{}.}
    \label{f:opt_effect}
\end{figure}

\subsection{Effectiveness of Environment Emulation} 
\label{sec:ev_diff_opt}

We evaluated whether our environment emulation (Section~\ref{subsec:emulating}) effectively enhances dynamic analysis by uncovering a broader set of system events from each container.\\

\noindent\textbf{Container Options.} We examined whether various Docker command options, as emulated by \ourtool{}, influence the privileges required by containers. The options tested were: (i) -{}-interactive, (ii) -{}-tty, (iii) -{}-detach, (iv) -{}-init, (v) -{}-network, (vi) -{}-publish-all, (vii) -{}-cpus, (viii) -{}-memory, (ix) -{}-hostname, and (x) -{}-stop-timeout, identified as popular choices based on our Docker Hub survey. For each run, we collected the system events generated by the container and calculated the number of syscalls invoked under two conditions: (i) without any options (i.e., \emph{baseline}) and (ii) with all tested options using \ourtool{}. Figure~\ref{f:opt_effect} illustrates the impact of these options on the number of syscalls triggered by 15 container images, selected for clarity from the 167 images tested. These 15 images invoked at least six additional distinct syscalls, demonstrating that \ourtool{} enables the discovery of approximately 11\% more syscalls. Across all 167 images, 51\% (85 images) invoked at least 5\% more syscalls when options were applied, and 6\% (10 images) required over 20\% additional capabilities with certain options.\\

\begin{figure}[t]
    \centering
    \includegraphics[width=\linewidth]{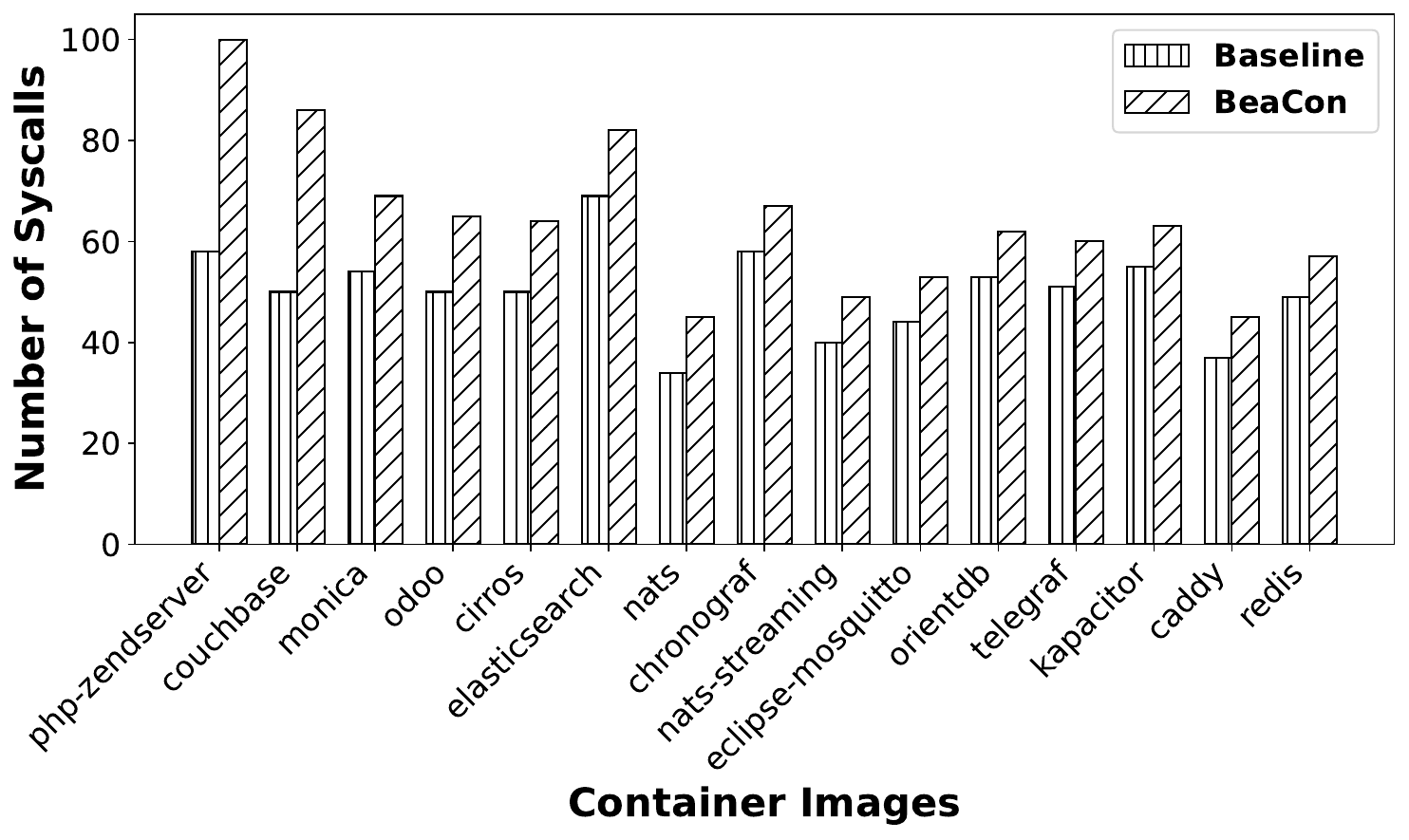}
    \caption{The number of syscalls observed without a workload (\emph{baseline}) compared to those recorded under workloads generated by \ourtool{}.}
    \label{f:workload_effect}   
\end{figure}

\noindent\textbf{Workloads.} We conducted a similar experiment to evaluate the effect of different workloads on the system events generated by containers during execution. For this purpose, we selected 71 official Docker container images compatible with the YCSB benchmarking tool. As in the previous experiment, we first ran each container and collected the system events generated without any workload applied (i.e., \emph{baseline}), then repeated the process while applying eight distinct workloads using \ourtool{} (as defined in Table~\ref{tab:Workload_Prop}). Figure~\ref{f:workload_effect} illustrates the number of syscalls recorded for 15 container images, selected for clarity from the 71 images tested. On average, these containers invoked 12 additional syscalls (a 22\% increase) when subjected to workloads generated by \ourtool{}. Across all 71 images, 39\% of containers (28 images) exhibited at least a 5\% increase in unique syscalls under varying workloads. Additionally, 4\% of containers (3 images) required 10\% more capabilities when workloads were applied.

\subsection{Emulation Space Reduction via Heuristics} 
\label{sec:ev_indp}

We evaluated the effectiveness of \ourtool{}'s heuristic in reducing the search space for large-scale emulation environments (Section~\ref{subsec:heuristics}). Specifically, we selected two Docker options (\texttt{-it} and \texttt{--publish-all}) and two workloads targeting \texttt{read} and \texttt{update} operations, denoted as $o_1$, $o_2$, $w_1$, and $w_2$, respectively. Among the available images, 64 official containers were compatible with the selected configurations. By pairing each container with six environments generated from all combinations of these options and workloads, we formed a total of 384 container–environment pairs.

\begin{figure*}[!t]
    \centering
    \includegraphics[width=.95\linewidth,clip,trim={1mm 1mm 1mm 1mm}]{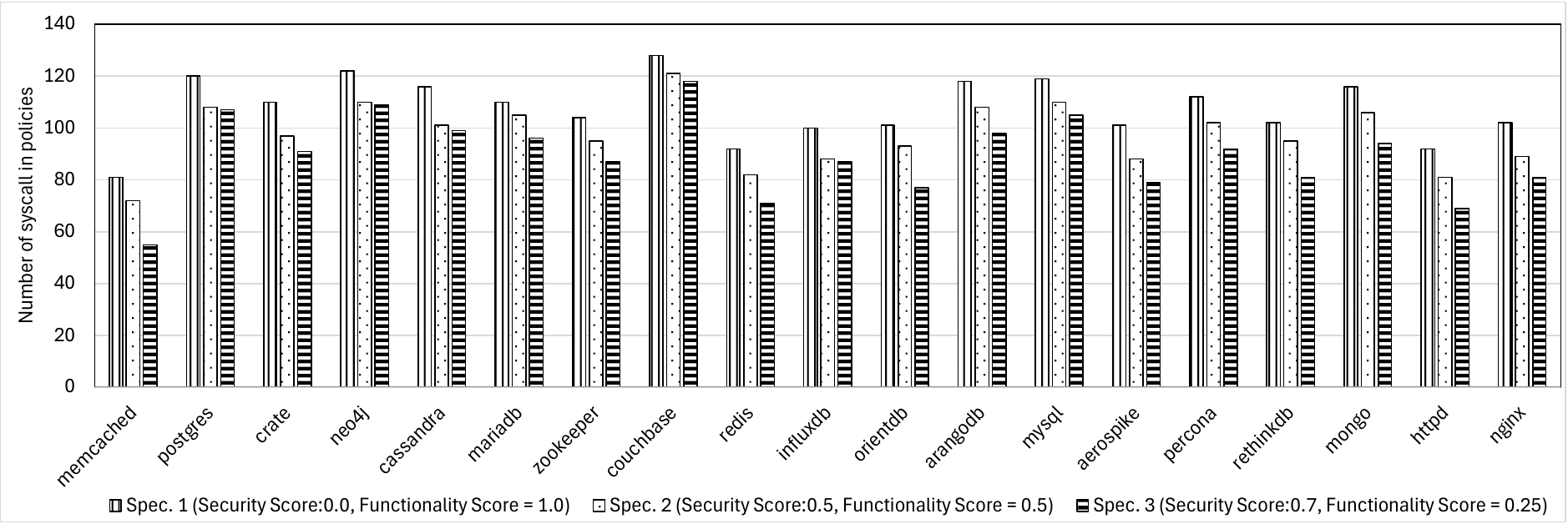}
    \caption{Policy size under tunable specifications. Spec. 1 (vertical stripped bars) aims to minimize container failures with a security score of 0 and a functionality score of 1. In contrast, Spec. 3 (horizontal stripped bars) prioritizes stricter security policies, reflected in its security score of 0.7, while its functionality score is 0.25. Spec 2 (spotted bars) targets balanced functionality and security (0.5, 0.5).}
    \label{f:workloadserial}
\end{figure*}

We monitored the system events generated when each Docker option or workload was applied individually ($E(e)$, where $e \in \{o_1, o_2, w_1, w_2\}$), and then again when two were applied together ($E(e_1, e_2)$, where $e_1, e_2 \in \{o_1, o_2, w_1, w_2\}$). To assess whether combined behavior can be inferred from individual cases, we compared the union of individual event sets ($E(e_1) \cup E(e_2)$) with the actual set observed from concurrent execution ($E(e_1, e_2)$). In 62.5\% of the 384 container–environment pairs, the two sets were identical ($E(e_1) \cup E(e_2) = E(e_1, e_2)$), and in the remaining 37.5\%, they differed by only one system call ($|E(e_1) \cup E(e_2)| - |E(e_1, e_2)| = 1$). Across all cases, there were no differences in requested capabilities. These results suggest that \ourtool{} can accurately infer the system event set of a complex environment by composing those from simpler configurations, reducing the need for exhaustive emulation and improving scalability.

\subsection{Performance of Event Monitoring}
\label{sec:ev_perf}

\begin{figure}[t]
    \centering
    \includegraphics[width=\linewidth,clip,trim={1mm 1mm 1mm 1mm}]{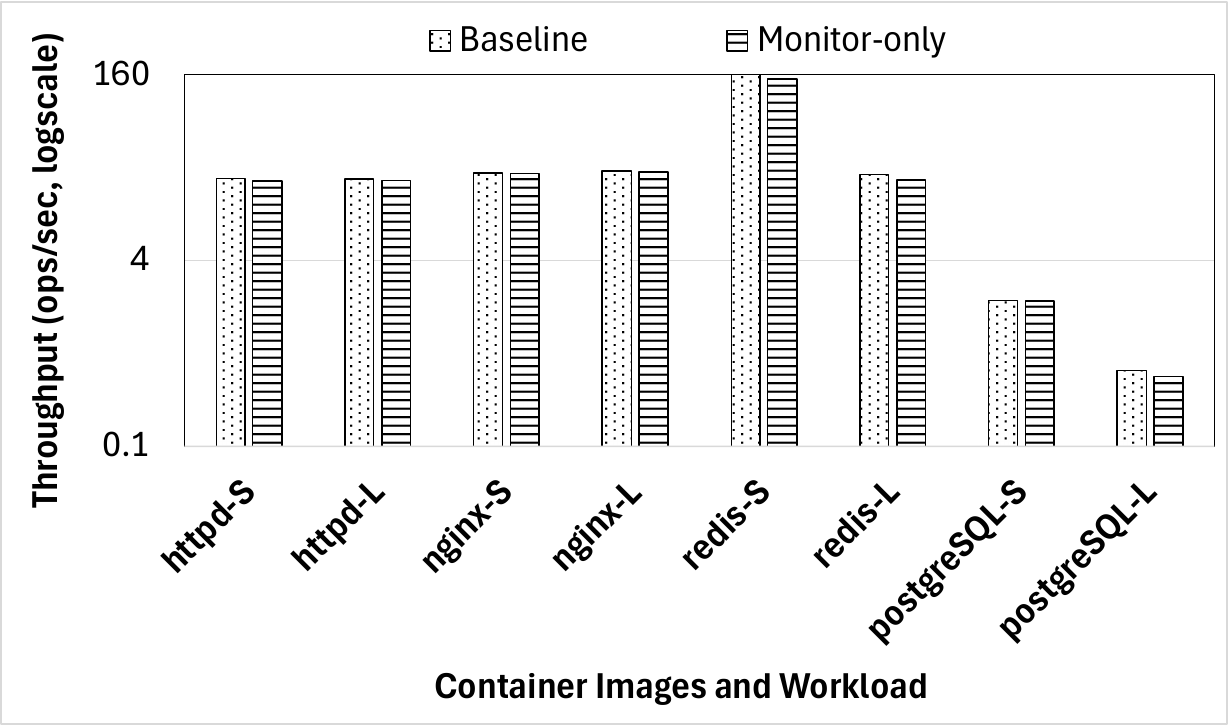}
    \caption{Performance of event monitoring. Request throughput (ops/sec, log scale) for two payload sizes per image (S=1 KB, L=64KB). Bars compare Baseline (without monitoring) and Monitor-only.}
    \label{fig:throughput}
\end{figure}

\begin{figure}[t]
    \centering
    \includegraphics[width=\linewidth,clip,trim={1mm 1mm 1mm 1mm}]{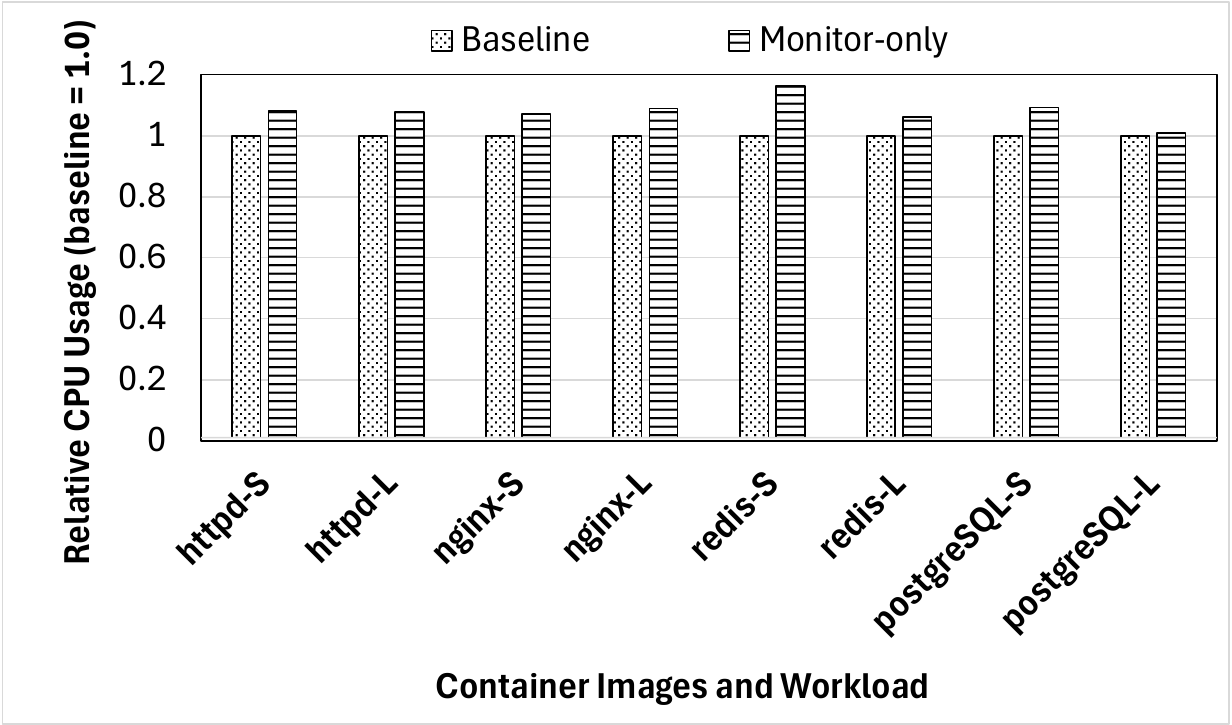}
    \caption{CPU cost of event monitoring (normalized). Relative CPU usage (baseline = 1.0) for the same scenarios as Fig.~\ref{fig:throughput}}
    \label{fig:cpu}
\end{figure}

We evaluated the cost of event monitoring by enabling only \ourtool{}'s eBPF pipeline (hooks on sys\_enter/cap\_capable, ring-buffer export, in-memory aggregation) and leaving the decision agent disabled. On two web servers (httpd, nginx) and two databases (Redis, PostgreSQL), each executed with two mixed workloads (Both 5000 operations of Read and Write, Different field length: S=1 KB, L=64KB). Monitoring agent causes a small, consistent throughput reduction by an average of 5.6\%, and relative CPU rises by 0.8-9\% (Figure~\ref{fig:throughput} and \ref{fig:cpu}). The log-scale throughput axis presents absolute differences between web and DB visible, but the relative impact of monitoring remains similar across images and payload sizes. This result indicates that the eBPF collection path adds modest overhead regardless of container types and workloads.

\subsection{Impact of Security and Functionality Scores}
\label{sec:ev_need_of_spec}

To evaluate the impact of varying security and functionality scores on generated policies (Section~\ref{sec:decision_agent}), we conducted a comparative analysis using three distinct specifications along the security-functionality axis: Spec-1 (prioritize functionality; target security 0.0, functionality 1.0), Spec-2 (balanced; 0.5, 0.5), and Spec-3 (prioritize security; 0.7, 0.25). Each specification is instantiated with the same scoring formulation, which loosens or tightens the CVSS-based exclusion and the required functionality threshold to meet the target scores. We evaluate 19 official images (17 databases + 2 web servers) selected from our corpus and generate environments using the nine Docker options/workload settings in Table~\ref{tab:Workload_Prop}.

Figure~\ref{f:workloadserial} reports the policy size (number of allowed syscalls) for each image under the three specifications. As the security target tightens from Spec-1 to Spec-3, policy size decreases monotonically for every image, i.e., stricter specifications yield smaller allowlists. The balanced Spec-2 consistently lands between the two extremes—shrinking policies relative to Spec-1 while retaining substantially more functionality than Spec-3 by design (0.5 vs. 0.25). In other words, \ourtool{}’s synthesis is tunable: operators are not forced to choose a fixed policy, but they can select a middle ground that improves security without the sharp functionality drop of the security-first extreme.

\subsection{Comparison with Static Analysis}
\label{sec:ev_CVE}

We evaluated whether \ourtool{} outperforms static analysis-based solutions—namely Chestnut~\cite{Chestnut}, Sysfilter~\cite{demarinis2020sysfilter}, and Confine~\cite{ghavamnia2020confine}—in reducing the attack surface.
Note that we exclude dynamic-analysis-based approaches such as Speaker~\cite{lei2017speaker} and Timeloops~\cite{Timeloops}, as they do not emulate external environments like \ourtool{}. Ensuring a fair comparison would require per-container workloads, which is a key contribution of \ourtool{}. We therefore focus our evaluation on widely adopted static-analysis-based solutions.
As a first step, we examined the applicability of these tools across all 167 container images in our dataset. Only 54 images were compatible with all solutions, so we restricted our evaluation to this subset. Our goal was to compare the number of syscalls permitted by each static analysis-based tool and by \ourtool{}. However, because Chestnut and Sysfilter are designed for analyzing standalone binaries rather than containerized environments, they were unable to directly identify the executables within containers. To address this limitation, we extracted all ELF binaries executed during container runtime and applied this list to both tools, thereby avoiding dependency issues caused by shared libraries. After generating the policies, we executed the containers with each policy in place to validate their correctness.

\begin{figure}[t]
    \centering
    \includegraphics[width=.9\linewidth, trim={0 0 0 1cm}]{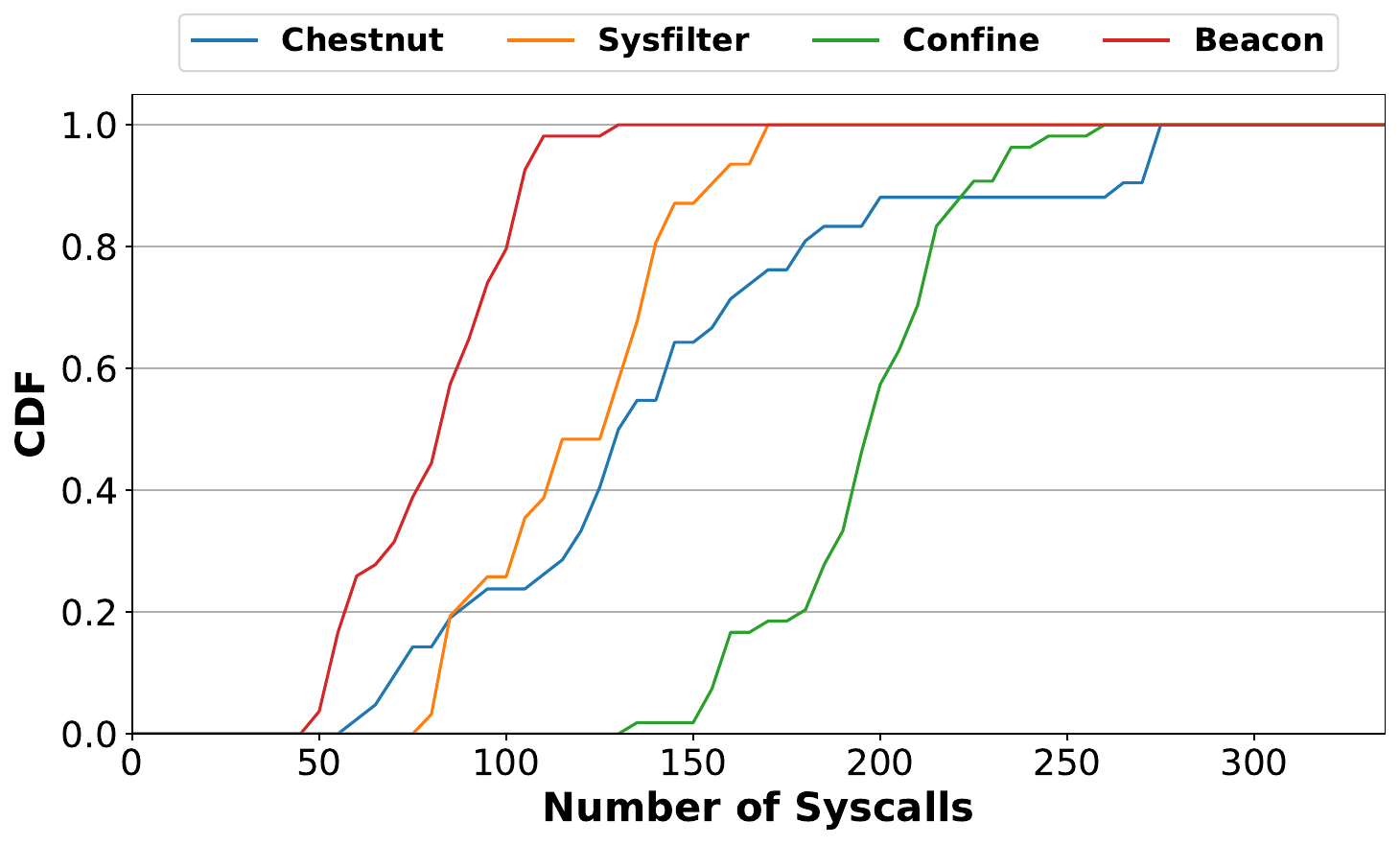}
    \caption{Comparison of the number of syscalls observed by previous solutions and \ourtool{}. A CDF curve located further to the right indicates a larger attack surface, as more syscalls are allowed.}
    \label{fig:prev_comparison}
\end{figure}

Figure~\ref{fig:prev_comparison} shows the cumulative distribution function (CDF) of the number of syscalls observed by each of the previous solutions and \ourtool{}. For a fair evaluation against static analysis approaches, we maximized the functionality score in order to include every syscall observed during environment emulation. Our experiments show that solutions using static analysis significantly overestimate container privileges, being Sysfilter the solution that produces tighter policies, and Confine the one that gives the most privileges to containers. In about 80\% of the containers, the policies that BeaCon creates contain +100 fewer syscalls than those generated by Confine.

\begin{table}[t!]
\centering
\caption{Security evaluation of Confine and \ourtool{} policies. Columns show CVEs, attack vectors, and the number of containers protected by each policy. \textit{Sec.} and \textit{Func.} refer to policies optimized for maximum security and functionality, respectively.}
\small
\resizebox{\linewidth}{!}{
\begin{tabular}{c c c c c c}
\toprule
\multirow{2}{*}{\textbf{CVEs}} & \multicolumn{2}{c}{\textbf{Attack Vectors}} & \multirow{2}{*}{\textbf{Confine}} & \multicolumn{2}{c}{\textbf{BeaCon}} \\ \cmidrule(l){2-3} \cmidrule(l){5-6} 
& \textbf{Capability} & \textbf{Syscalls} & & \textit{\textbf{Sec.}} & \textit{\textbf{Func.}} \\ \midrule
CVE-2017-7308       & NET\_RAW     & socket                &   0/71 & 70/71 & 70/71 \\
CVE-2017-5123       & \textit{n/a}      & waitid                & 60/71 & 71/71 & 71/71 \\
CVE-2016-8655       & NET\_RAW     & socket, setsockopt    &   4/71 & 70/71 & 70/71 \\
CVE-2016-0728       & \textit{n/a}      & keyctl                & 71/71 & 71/71 & 71/71 \\
CVE-2014-9529       & \textit{n/a}      & keyctl                & 71/71 & 71/71 & 71/71 \\
CVE-2015-8660       & SYS\_ADMIN   & mount                 &  3/71 & 67/71 & 70/71 \\
CVE-2016-5195       & \textit{n/a}      & madvise               &   0/71 & 22/71 & 28/71 \\
CVE-2019-11815      & \textit{n/a}      & clone, unshare        & 65/71 & 71/71 & 71/71 \\
CVE-2013-1959       & \textit{n/a}      & write                 &   0/71 &  0/71 &  0/71 \\
CVE-2015-8543       & NET\_RAW     & socket                &   0/71 & 70/71 & 70/71 \\
CVE-2017-17712      & NET\_RAW     & sendto, sendmsg       &   0/71 & 70/71 & 70/71 \\
CVE-2018-14634      & \textit{n/a}      & waitid                & 60/71 & 71/71 & 71/71 \\
CVE-2017-14954      & \textit{n/a}      & waitid                & 60/71 & 71/71 & 71/71 \\
CVE-2014-5207       & SYS\_ADMIN   & mount                 &  3/71 & 67/71 & 70/71 \\
CVE-2018-12233      & \textit{n/a}      & setxattr              &  10/71 & 70/71 & 70/71 \\
CVE-2019-13272      & PTRACE       & ptrace                & 71/71 & 71/71 & 71/71 \\
CVE-2018-1000199    & PTRACE            & ptrace                &   71/71 & 71/71 & 71/71 \\
CVE-2014-4699       & \textit{n/a}      & fork, clone, ptrace   &   70/71 & 71/71 & 71/71 \\
CVE-2014-7970       & SYS\_ADMIN        & pivot\_root           &   70/71 & 67/71 & 71/71 \\
CVE-2019-10125      & \textit{n/a}      & io\_submit            &   71/71 & 71/71 & 71/71 \\
CVE-2017-6001       & \textit{n/a}      & perf\_event\_open     &   71/71 & 71/71 & 71/71 \\
CVE-2016-2383       & SYS\_ADMIN        & bpf                   &   71/71 & 71/71 & 71/71 \\
CVE-2018-11508      & \textit{n/a}      & adjtimex              &   64/71 & 71/71 & 71/71 \\
CVE-2014-8133       & \textit{n/a}      & set\_thread\_area     &   71/71 & 71/71 & 71/71 \\
CVE-2017-11176      & \textit{n/a}      & mq\_notify            &   71/71 & 71/71 & 71/71 \\
CVE-2014-9903       & \textit{n/a}      & sched\_getattr        &   71/71 & 71/71 & 71/71 \\
CVE-2010-3066       & \textit{n/a}      & io\_submit            &   71/71 & 71/71 & 71/71 \\
CVE-2011-1182       & \textit{n/a}      & rt\_(tg)sigqueueinfo  &   71/71 & 71/71 & 71/71 \\
CVE-2018-13053      & \textit{n/a}      & clock\_nanosleep      &   69/71 & 71/71 & 71/71 \\
CVE-2016-7911       & \textit{n/a}      & ioprio\_get           &   71/71 & 71/71 & 71/71 \\
CVE-2017-14954      & \textit{n/a}      & waitid                &   60/71 & 71/71 & 71/71 \\
CVE-2017-5123       & \textit{n/a}      & waitid                &   60/71 & 71/71 & 71/71 \\
CVE-2010-4250       & \textit{n/a}      & inotify\_init1        &   45/71 & 71/71 & 71/71 \\
CVE-2010-4083       & \textit{n/a}      & semctl                &   54/71 & 67/71 & 68/71 \\
CVE-2019-9857       & \textit{n/a}      & inotify\_add\_watch   &   45/71 & 71/71 & 71/71 \\
CVE-2009-0859       & \textit{n/a}      & shmctl                &   41/71 & 71/71 & 71/71 \\
CVE-2010-4072       & \textit{n/a}      & shmctl                &   41/71 & 71/71 & 71/71 \\
CVE-2015-7613       & \textit{n/a}      & (sem,msg,shm)get      &   63/71 & 66/71 & 71/71 \\
CVE-2009-1961       & \textit{n/a}      & splice                &   24/71 & 71/71 & 71/71 \\
\rowcolor{gray!20}
CVE-2012-3375       & \textit{n/a}      & epoll\_ctl            &   34/71 & 71/71 & 71/71 \\
CVE-2016-4997       & \textit{n/a}      & setsockopt            &   4/71  &  7/71 &  8/71 \\
CVE-2010-2478       & \textit{n/a}      & ioctl                 &   0/71  &  0/71 &  5/71 \\
CVE-2009-0745       & \textit{n/a}      & ioctl                 &   0/71  &  0/71 &  5/71 \\
\rowcolor{gray!20}
CVE-2012-3511       & \textit{n/a}      & madvise               &    0/71 & 22/71 & 28/71 \\
CVE-2017-18208      & \textit{n/a}      & madvise               &    0/71 & 22/71 & 28/71 \\
\midrule
\textbf{Average Mitigation Ratio} & & & 60.5\% & 85.9\% & 60.47\% \\
\bottomrule
\end{tabular}
}
\label{tab:secEffect}
\end{table}

\subsection{Mitigation of Known Vulnerabilities}
\label{sec:known_attacks}

To evaluate the security effectiveness of \ourtool{}, we conducted an analysis to measure how \ourtool{} reduces the attack surface of the host OS kernel. This measurement was performed by calculating the ratio of mitigated containers among all container images examined. Table~\ref{tab:secEffect} shows the result of our analysis on container images with respect to CVEs. The first column lists the CVE identifiers, while the second and third column describe the capabilities and syscalls identified as attack vectors associated with each CVE. To identify relevant attack vectors, we manually collected and interpreted documentation on attack scenarios. We then examined the security policies generated by Confine~\cite{ghavamnia2020confine} and \ourtool{} to determine whether these vulnerable system events were allowed. 

We generated security policies for 71 images\footnote{We selected those 71 images because they can run in a headless manner, which is essential for automatic policy generation in \ourtool{}. Consequently, we exclude images that require manual setup, credentials, or an orchestrator, most of which were used in the Confine experiment.} which were subjected to workloads using YCSB, enabling us to evaluate the security effectiveness when considering external environments using both \ourtool{} and Confine.
The columns titled ``Confine'' and ``\ourtool{}'' in Table~\ref{tab:secEffect} represent the number of container images secured by each mechanism out of the total analyzed. Because \ourtool{}'s security policies depend on balancing security and functionality, the number of secured containers is shown separately. The ``\emph{Sec.}'' column indicates the number of secured containers when prioritizing security, while the ``\emph{Func.}'' column represents the number secured when maximizing functionality.

\ourtool{} demonstrates superior performance in terms of securing containers. For example, in the case of CVE-2012-3375, Confine prevents the vulnerability in 44\% of the containers, which represents the best performance achieved by Confine. However, \ourtool{} outperforms this by successfully preventing the vulnerability in all 71 containers. On the other hand, when dealing with CVE-2012-3511, which represents the worst performance of \ourtool{}, it still secures 31\% to 39\% of the containers, while Confine prevents only one. Overall, Confine achieves a 58.6\% mitigation ratio on average, while \ourtool{} achieves 85.3\%. These results highlight \ourtool{}'s ability to considerably reduce the attack surface and mitigate realistic threats more effectively than Confine. Furthermore, even the ``\emph{Func.}'' policy of \ourtool{}, which allows more syscalls than the ``\emph{Sec.}'' policy, can protect containers better than Confine.

\subsection{Security Case Studies}
\label{sec:securityusecases}

Although the idea of blocking specific syscalls to prevent container exploits is well established, we present two case studies demonstrating how \ourtool{} selectively enforces such restriction based on its environment-aware dynamic profiling. Unlike prior solutions such as Confine~\cite{ghavamnia2020confine} and Speaker~\cite{lei2017speaker}, which often allow or deny syscalls indiscriminately, \ourtool{} removes risky sycalls or capability only when they are not required for observed functionality. This describes how \ourtool{} achieves strong security outcomes while preserving availability.

\subsubsection{(CVE-2016-5195) Dirty CoW}

Dirty CoW~\cite{cve-2016-5195} is a widely known and serious vulnerability that affected all Linux-based operating systems in 2018. It abuses the Copy-On-Write (CoW) mechanism\footnote{The CoW is a technique for efficient memory management; the Linux kernel creates the copy only when a write operation is requested.} in the Linux kernel to turn a read-only mapping into a writable one. In general, when performing a write operation on read-only data, the Linux kernel creates a private copy on memory and the write operation is executed in the latter. However, it has been found that adversaries can exploit a race condition to write arbitrary contents into the read-only data. In particular, the core idea behind this attack is to create two threads that invoke the \texttt{madvise} and \texttt{write} syscalls, respectively, with the address of memory mapping, where the first thread is responsible for repeatedly asking the Linux kernel to free the private copy while the latter thread attempts to write data by invoking the \texttt{write} syscall. The host on which the containers run uses the Linux kernel v4.4.0 (Ubuntu 16.04).\\

\begin{figure}[t]
    \centering
    \subfloat[Attack succeeds with Confine's policy.]{
        \includegraphics[width=.9\linewidth]{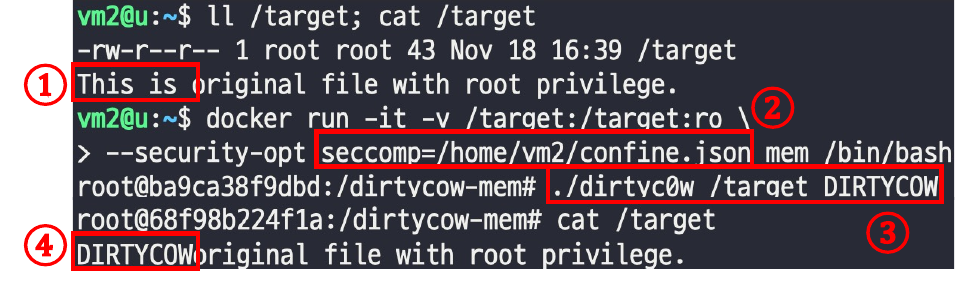}
        \label{f:dirtycow_file_a}
    }
    
    \subfloat[Attack is blocked with \ourtool{}'s policy.]{
        \includegraphics[width=.9\linewidth]{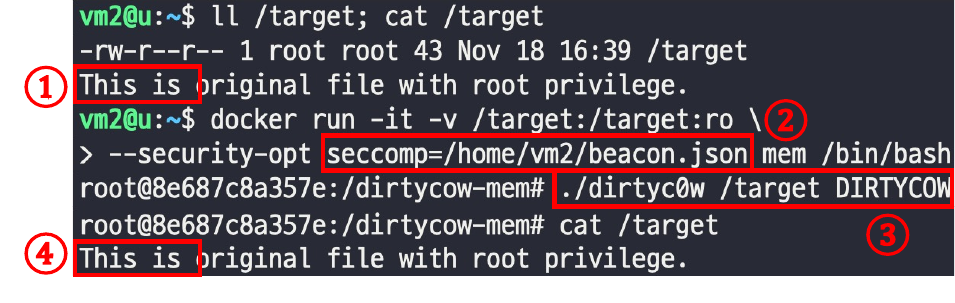}
        \label{f:dirtycow_file_b}
    }
    \caption{File modification attack using Dirty CoW: \protect\circled{1} the original target file content, \protect\circled{2} Seccomp policies generated by Confine and \ourtool{}, \protect\circled{3} attack execution, and \protect\circled{4} the target file content after the attack.}
\end{figure}

\noindent\textbf{File Modification Attack.} By default, containers are not allowed to access files in the host. In practice, some containers require the host to share its files (e.g., Postgres, composer, Redmine and PHP containers need to share the system file \texttt{/etc/passwd} for user authentication) by mounting a host volume into the container's file system. Although these host files are typically shared with read-only permission, a malicious container that leverages the Dirty CoW vulnerability could gain writing privileges in the host-shared file.

To test the attack, we reproduced the Dirty CoW’s Proof of Concept (PoC)~\cite {DirtyCowPoC} and observed that it requires the invocation of the following six syscalls: \texttt{mmap}, \texttt{madvise}, \texttt{write}, \texttt{open}, \texttt{lseek}, \texttt{fstat}. As \texttt{mmap}, \texttt{write}, \texttt{open}, \texttt{lseek}, \texttt{fstat} are popular syscalls, we focused on measuring how often \ourtool{} includes the \texttt{madvise} syscall in the policies it generates for containers. 

\ourtool{} excludes \texttt{madvise} syscall from the policy for 22 out of 71 official server containers that never invoke \texttt{madvise} syscall regardless of their environment as described in Figure~\ref{f:dirtycow_file_b}. Consequently, any attemp to invoke \texttt{madvise} at runtime would be blocked by Seccomp. On the other hand, \texttt{madvise} is one of the syscalls allowed within Docker's default policy, it will fail to prevent DirtyCow. We found that Confine also fails to block \texttt{madvise} syscall as shown in Figure~\ref{f:dirtycow_file_a} in all 71 containers except for the \emph{nat-streaming}~\cite{nats-streaming}. In 6 containers \texttt{madvise} appears only under specific environments where \ourtool{} is adjustable: they include the syscall for improved functionality or exclude it to enhance security.

\subsubsection{(CVE-2020-14386) Raw-socket} \label{subsection72}

CVE-2020-14386~\cite{cve-2020-14386} exploits the packet handler in the Linux kernel by abusing the \texttt{socket()} and \texttt{setsockopt()} syscalls to inject an abnormal value into a raw-socket packet handler. An adversary can manipulate the ring buffer headroom size (i.e., \texttt{PACKET\_RESERVE}), which will affect the way packet headers are parsed within the function. This causes overflow, resulting in memory corruption when accessing the remaining part of the packet. Note that an adversary can abuse this vulnerability in most containers since many container platforms typically allow to use of a raw-socket with the \texttt{CAP\_NET\_RAW} capability (e.g., Docker).\\

\begin{figure}[t]
    \centering
    \subfloat[Attack succeeds with Confine and Speaker policies.]{
        \includegraphics[width=0.85\linewidth]{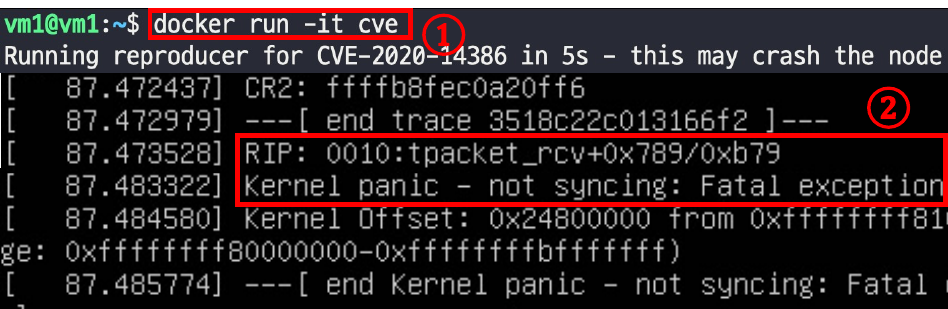}
        \label{f:afpacket-suc}
    }
    
    \subfloat[Attack is blocked with \ourtool{}'s policy.]{
        \includegraphics[width=.87\linewidth]{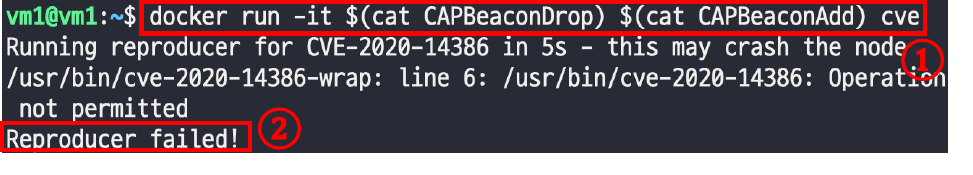}
        \label{f:afpacket-fail}
    }
    \caption{Kernel panic attack exploiting the raw socket vulnerability: \protect\circled{1} attack execution and \protect\circled{2} system response.}
\end{figure}

\noindent\textbf{Kernel Panic Attack.} Adversaries within containers can exploit this vulnerability by injecting an arbitrary address into the raw-socket handler, causing the kernel to enter an unexpected state. In our experiments, we utilized a proof-of-concept (PoC) exploit~\cite{poc_cv2-2020-14386} targeting Linux kernel version 5.4.0 (Ubuntu 20.04). Since many containers legitimately invoke the \texttt{socket} and \texttt{setsockopt} system calls, tools such as Confine and Speaker—which generate security policies based solely on syscalls—are unable to block this type of attack. In contrast, \ourtool{} successfully mitigates the threat by removing the \texttt{CAP\_NET\_RAW} capability from the container during policy generation. As shown in Figures~\ref{f:afpacket-suc} and \ref{f:afpacket-fail}, the syscall-only policies generated by Confine and Speaker fail to prevent the attack, while the combined syscall and capability-based policy produced by \ourtool{} effectively blocks it.

\section{Discussion and Limitations}
\label{sec:discussion}
In this section, we discuss the design considerations and limitations of \ourtool{}.

\noindent\textbf{Dependency to Workloads.} The efficacy of \ourtool{} hinges upon the workloads provided by cloud providers, a common challenge to all dynamic analysis-based solutions. While utilizing realistic workloads sourced from cloud environments would be optimal, it often proves unfeasible due to the difficulty of obtaining such data in a container-agnostic manner (realistic workloads data is typically not publicly available) for our experiments. As a mitigation strategy, we leverage YCSB, a versatile benchmarking tool whose parameters and profiling time can be fine-tuned to simulate diverse workloads. However, given the general-purpose nature of \ourtool{}, it remains plausible to devise tailored policies aligned with cloud providers' requirements by leveraging alternative benchmarking tools or real-world workloads.\\

\noindent\textbf{Expanding Policy Generation to Include Other LSMs.} Linux Security Modules (LSMs) such as AppArmor~\cite{AppArmor} and SELinux~\cite{SELinux} offer various security models, including mandatory access control (MAC) for system resources like the file system. While these LSMs and Seccomp filters are valuable tools for enhancing the Linux system security, their primary approaches are fundamentally different: Seccomp filters focus on controlling syscalls, whereas LSMs primarily target file-based access control. Despite the distinct focus of \ourtool{} on automatically generating security policies based on dynamic analysis, it is crucial to recognize the complementary nature of LSMs and our approach. \ourtool{} and LSMs operate orthogonally, with each addressing different aspects of system security. While \ourtool{} primarily focuses on generating policies for Seccomp filters, there is potential for expansion to include support for generating policies compatible with other LSMs.\\

\noindent\noindent\textbf{Including Syscall Parameters in Policy Generation.} Cloud providers have the capability to filter syscalls not only based on the syscall itself but also on its parameters, enabling the implementation of more stringent security policies. However, \ourtool{}'s design choice does not incorporate syscall parameter filtering. This decision stems from the fact that parameters such as process identifiers or file descriptors could vary randomly with each container execution, resulting inconsistent policies. While it is possible to improve by classifying the types of resources pointed by each parameter, \ourtool{} prioritizes efficiency in guiding the balance between security and functionality. As illustrated in Figure~\ref{f:workloadserial}, even without explicit consideration of syscall parameters, \ourtool{} demonstrates significant potential adjustments, highlighting the effectiveness of our approach. Thus, while parameter consideration is important, it falls outside the scope of our current work.\\

\noindent\textbf{Balancing Detection Precision and Recall.} Auto-generated policies often involve a trade-off between detection precision and recall. A stricter policy (high precision) may disrupt container operations by blocking essential system events, whereas a more permissive policy (high recall) risks allowing critical system events that could be exploited. Evaluating precision and recall in this context, however, is inherently problematic because the ground truth is defined by the cloud provider’s security posture and priorities. If a provider prioritizes security, \ourtool{} must minimize the set of allowed system events, even at the expense of recall. Given this variability, evaluating precision and recall as fixed metrics is not meaningful for this work. Instead, we propose new metrics—security and functionality scores—that dynamically adjust container privileges to align with the cloud provider's specific requirements. These scores provide a flexible framework for tailoring policies to balance security and operational needs, rather than adhering to static precision and recall evaluations.\\

\noindent\textbf{Handling Zero-Day Vulnerabilities.}
\ourtool{} relies on the CVSS framework to measure the security scores of containers. This reliance means that if a container uses syscalls or capabilities associated with zero-day vulnerabilities, for which CVSS scores are unavailable, accurate security scoring becomes infeasible. However, the primary objective of \ourtool{} is to reduce the container's attack surface by restricting privileges, not to directly detect or defend against specific zero-day attacks. While CVSS provides a reliable metric for evaluating the severity of known vulnerabilities, addressing zero-day vulnerabilities lies beyond the scope of this work. Attempting to assess zero-day vulnerabilities without CVSS risks inaccurately estimating their impact, potentially compromising the reliability of security evaluations.

\begin{table*}[t]
\scriptsize

\caption{Comparison between \ourtool{} and other state-of-the-art attack surface reduction solutions. \fullcercle~indicates that the solution satisfies the requirement and \emptycercle~is used when the requirement is not fulfilled.}
\centering

\begin{tabular}{c c P{1.5cm} P{1.5cm} P{1.5cm} P{1.5cm} P{1.5cm}}
\toprule
\textbf{Solutions} & \textbf{Approach} & \textbf{Container-specific} & \textbf{Syscall-aware} & \textbf{Capability-aware} & \textbf{Environment-aware} & \textbf{Policy-adjustable} \\ \midrule
Confine~\cite{ghavamnia2020confine} & Static & \fullcercle & \fullcercle & \emptycercle & \emptycercle & \emptycercle \\ \midrule
Chestnut~\cite{Chestnut} & Static & \emptycercle & \fullcercle & \emptycercle &  \emptycercle &  \emptycercle \\ \midrule
Sysfilter~\cite{demarinis2020sysfilter} & Static & \emptycercle & \fullcercle & \emptycercle & \emptycercle & \emptycercle \\ \midrule

Speaker~\cite{lei2017speaker} & Dynamic & \fullcercle & \fullcercle &\emptycercle & \emptycercle & \emptycercle \\ \midrule
Timeloops~\cite{Timeloops} & Dynamic  &  \fullcercle & \fullcercle & \emptycercle &  \emptycercle &  \emptycercle \\ \midrule
Podman~\cite{podman} & Dynamic &  \fullcercle & \fullcercle & \emptycercle &  \emptycercle &  \emptycercle \\ \midrule
Sysdig~\cite{sysdig} & Dynamic &  \fullcercle & \fullcercle & \emptycercle &  \emptycercle &  \emptycercle \\ \midrule
Decap~\cite{hasan2022decap} & Static \& Dynamic & \emptycercle & \emptycercle & \fullcercle & \emptycercle & \emptycercle \\ \midrule
Sysverify~\cite{zhan2022shrinking} & Static \& Dynamic & \emptycercle & \fullcercle & \emptycercle  & \emptycercle  & \emptycercle  \\ \midrule
µPolicyCraft~\cite{blair2023automated} & Static \& Dynamic & \fullcercle  & \fullcercle & \emptycercle  & \fullcercle  & \emptycercle  \\ \midrule

\textbf{\ourtool{} (our work)}  & Dynamic  &  \fullcercle & \fullcercle & \fullcercle &  \fullcercle &  \fullcercle \\

\bottomrule
\end{tabular}
\label{tab:comparison}
\end{table*}

\section{Related Work}
\label{sec:related_work}

In this section, we present existing solutions\footnote{It is important to note that we also incorporate solutions that target general Linux environments while our focus is solely on containers.} for attack surface reduction based on the usage of dynamic and static analysis to automatically retrieve the privileges used by an application. Next, we conduct an in-depth analysis of the existing solutions and reveal their limitations to motivate our work. We refer to Table~\ref{tab:comparison} for the comparison.\\

\noindent\textbf{Dynamic Analysis-based Solutions.}
Speaker~\cite{lei2017speaker} and Timeloops~\cite{Timeloops} are the two state-of-the-art solutions that rely on dynamic analysis to obtain the privileges needed by containers automatically. In the profiling phase of a container, Speaker creates two policies, one for the booting phase (which takes roughly 2 minutes) and one for the running phase. This is because behavior of containers in these two phases are diverse and therefore require different privilege. With Speaker, containers are first run with the profile for the booting phase (which is typically more demanding in terms of required privileges), and after the booting phase is completed, the profile is updated and those syscalls no longer needed are removed. On the contrary, Timeloops takes a completely different approach that is inspired on the principle of least privileges. The authors propose to always start with an empty permissions list. This way, every time the container invokes syscalls that are not included in the policy, Timeloops evaluates whether the syscall can trigger vulnerabilities, and based on that, decides whether to add it to the policy. 

Although these solutions are promising, they have notable limitations. Speaker profiles containers without considering the environments they may encounter during actual execution, such as the influence of Docker parameters and workloads on container behavior and the syscalls invoked. This oversight can result in policies that cause container failures. Similarly, Timeloops lacks mechanisms to actively analyze external factors that trigger behavioral variations, leading to prolonged efforts to develop stable, error-free policies. While tools like Podman~\cite{podman} and Sysdig~\cite{sysdig} support dynamic syscall profiling, they do not incorporate profiling capabilities to address known vulnerabilities, which are critical for robust security (see Sections~\ref{ref_motivation} and \ref{sec:securityusecases}).\\

\noindent\textbf{Static Analysis-based Solutions.}
Other solutions apply static analysis techniques to discover the privileges needed by applications. These solutions can be divided into several categories depending on whether they require access to the source code or the binary and whether they work for any application or were designed for containers.

Confine~\cite{ghavamnia2020confine} is the state-of-the-art solution in applying static analysis techniques to containers to find the privileges they need. From the source code of the containers, Confine identifies all of its processes and library functions that they import, assigns library functions to system calls, and retrieves direct system call invocations from previously identified processes and libraries. In a parallel line of research, multiple static-based generic solutions have been proposed~\cite{Chestnut, demarinis2020sysfilter}. These solutions aim to automatically determine the privileges of generic applications by creating a call graph that describes the possible control flows within the application. Chestnut~\cite{Chestnut} offers a quick analysis of both source code and binaries, generating Seccomp policies. Sysfilter~\cite{demarinis2020sysfilter} filters system calls on any Linux binary to control which invocations are allowed.

Their main limitation is that identifying control flows from the container source code or binary is extremely hard, and profiles without knowledge of control flow are mostly overestimated. Also, some of these solutions are incompatible with some container images. For example, Kim et al.~\cite{9582167} reported that it is difficult for Confine~\cite{ghavamnia2020confine} to parse binaries that do not use the \emph{libc} library for syscall invocations. Similarly, Sysfilter~\cite{demarinis2020sysfilter} does not support containers that rely on Alpine as the underlying operating system while Chesnut cannot test all processes created inside a container.\\

\noindent\textbf{Hybrid Solutions.}
Recently, several solutions have adopted both techniques to generate more fine-grained security policies.
The majority of these approaches initially employ static analysis to extract syscalls, followed by a verification process based on dynamic analysis to eliminate overestimated ones.
Decap~\cite{hasan2022decap} performs an analysis of Linux program source code to construct syscall-to-capability mappings, which are refined via dynamic analysis. The mappings are utilized for finding minimum capabilities of Linux programs. 
Sysverify~\cite{zhan2022shrinking}, for example, leverages static analysis to obtain an overestimated syscall list and prunes it by utilizing dynamic analysis-based verification. μPolicyCraft~\cite{blair2023automated} takes micro execution---a similar concept with symbolic execution---to generate a container's effect graph that describes how it is likely to behave. The graph is then utilized for inspecting violations of running containers, comparing it with container telemetry.

Although these hybrid solutions exhibit enhanced effectiveness compared to static analysis-based methods, they still necessitate access to the source code of containerized applications, which constitutes an impractical assumption within cloud environments.
Furthermore, hybrid solutions still concentrate on reducing the attack surface by focusing solely on syscalls, thereby overlooking other critical factors such as container-specific environments, capabilities, and the intentions of cloud providers/tenants.
In contrast, \ourtool{} endeavors to address all these requirements, as shown in Table~\ref{tab:comparison}.

\section{Conclusion}
\label{sec:conclusion}

This paper presents the design and implementation of a novel approach for the automated generation of intention awareness policies suitable for containers. It is based on running containers and capturing the system events they request using dynamic analysis coupled with realistic environments. \ourtool{} is lightweight, non-intrusive, and compatible with any server type of container image. Our security use-cases demonstrate that \ourtool{} efficiently reduces the attack surface, thus preventing attacks from adversaries who exploit excessively granted privileges. Additionally, our evaluation results indicate that it is crucial to consider the environments under which containers are executed in order to assign only the minimum privileges required to containers.

% \section{Acknowledgments}

% We thank the anonymous reviewers for their insightful feedback, which helped improve this paper. This research was supported by Horizon Europe through the FLUIDOS project (GA#101070473), and by the UNICO I+D Cloud program funded by the Ministry of Economic Affairs and Digital Transformation and the European Union–NextGenerationEU as part of the Plan de Recuperación, Transformación y Resiliencia (PRTR) through the CLOUDLESS project. 
% \eduard{Eduard: Please do not modify the second sentence, as I need it to remain unchanged in order to properly claim that this work was carried out under these projects}

\printcredits

\bibliographystyle{unsrt}

\bibliography{references}

\clearpage

\bio{author_photo/bio_haney}
\textbf{Haney Kang} is a Ph.D. candidate in the School of Electrical Engineering at KAIST. He received both his B.S. and M.S. degrees from the same institution. His research focuses on security and high-performance networking systems for cloud environments, with particular interest in the interplay between systems security, and programming languages. He aims to build practical, scalable systems that strengthen cloud-native infrastructure while preserving efficiency.
\endbio

\vspace{0.7in}

\bio{author_photo/bio_marin}
\textbf{Eduard Marin} is a Senior Research Scientist and Team Lead at Telefónica Research, Spain. He received his Ph.D. from KU Leuven, Belgium, and his B.S. and M.S. degrees in Telecommunications Engineering from UPC, Spain. Following his Ph.D., he was a visiting researcher at the University of Padua (Italy) and a postdoctoral researcher at the University of Birmingham (UK). His research focuses on the intersection of networks, cloud computing and security \& privacy.
\endbio

\vspace{0.7in}

\bio{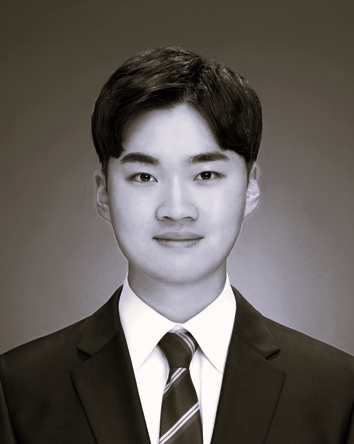}
\textbf{Myoungsung You} is an Assistant Professor in School of Electrical and Computer Engineering at University of Seoul.
He received his Ph.D. from the School of Electrical Engineering at KAIST, his M.S. degree from the Graduate School of Information Security at KAIST, and his B.S. degree from Chungbuk National University in Computer Engineering. His research interests include programmable network data planes, cloud security, and distributed systems.
\endbio

\vspace{0.7in}

\bio{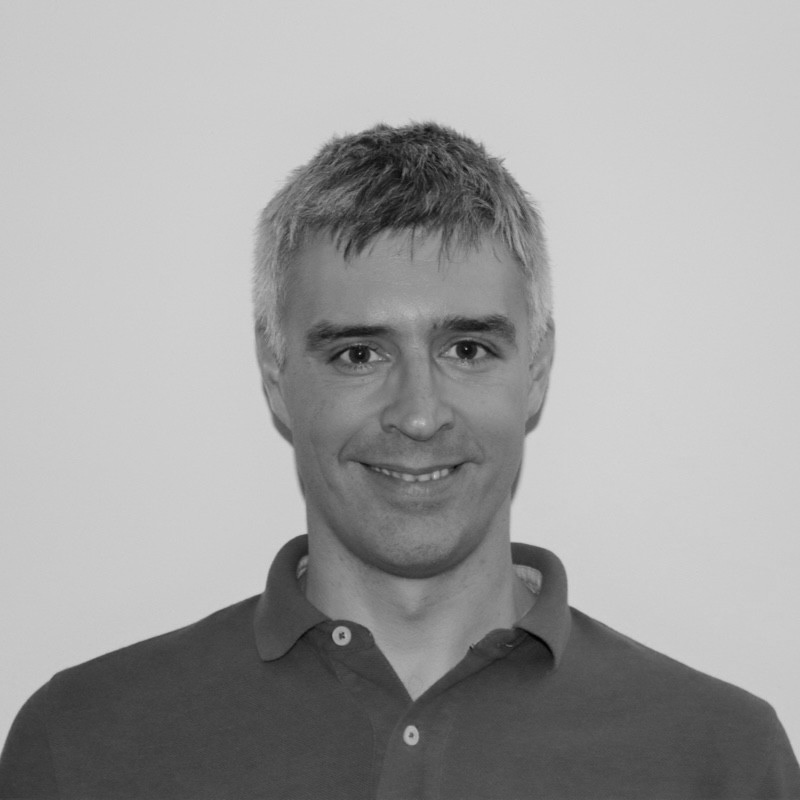}
\textbf{Diego Perino} is an organization manager, technical leader and scientist with passion to work in cutting edge projects with industrial impact. He currently is the director of the AI Institute of the Barcelona Supercomputing Center and he has been working for different companies in the ITC sector (Telefonica, Bell Labs, Orange Labs) and covered different technologies and research areas (above all AI, networks, systems). Apart from his industrial experience, he has also been very active in the scientific community with several publications, participation in conference committees, and editorial board contributions. He holds a Ph.D. from the Paris Diderot-Paris 7 University, MSc. from Politecnico di Torino, Eurecom Institute and Université de Nice-Sophia Antipolis.

\endbio

\vspace{0.7in}

\bio{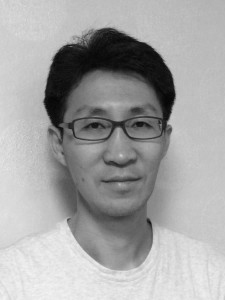}
\textbf{Seungwon Shin} is an Associate Professor in the School of Electrical Engineering at KAIST and an Executive Vice President at Samsung Electronics. He received his Ph.D. in Computer Engineering from Texas A\&M University, and his M.S. and B.S. degrees from KAIST, all in Electrical and Computer Engineering. His research interests include software-defined networking security, Dark Web analysis, and cyber threat intelligence.
\endbio

\vspace{0.7in}

\bio{author_photo/bio_kim}
\textbf{Jinwoo Kim} is an Assistant Professor in the School of Software at Kwangwoon University, Seoul, Republic of Korea. He received his Ph.D. from the School of Electrical Engineering at KAIST, his M.S. degree from the Graduate School of Information Security at KAIST, and his B.S. degree from Chungnam National University in Computer Science and Engineering. His research focuses on investigating security issues in software-defined networks and cloud systems.
\endbio

\end{document}